\documentclass[twocolumn]{aastex63}
\usepackage{amsmath}
\usepackage{color}
\def\beq{\begin{eqnarray}}
\def\eeq{\end{eqnarray}}

\begin{document}

\title{Contribution of neutrino-dominated accretion flows to cosmic MeV neutrino background}

\correspondingauthor{Tong Liu}
\email{tongliu@xmu.edu.cn}

\author[0000-0002-9130-2586]{Yun-Feng Wei}
\affiliation{Department of Astronomy, Xiamen University, Xiamen, Fujian 361005, China}

\author[0000-0001-8678-6291]{Tong Liu}
\affiliation{Department of Astronomy, Xiamen University, Xiamen, Fujian 361005, China}

\author[0000-0001-8390-9962]{Cui-Ying Song}
\affiliation{Physics Department and Tsinghua Center for Astrophysics, Tsinghua University, Beijing 100084, China}

\begin{abstract}
Neutrino-dominated accretion flows (NDAFs) are one of the important MeV neutrino sources and significantly contribute to the cosmic diffuse neutrino background. In this paper, we investigate the spectrum of diffuse NDAF neutrino background (DNNB) by fully considering the effects of the progenitor properties and initial explosion energies based on core-collapse supernova (CCSN) simulations, and estimate the detectable event rate by Super-Kamiokande detector. We find that the predicted background neutrino flux is mainly determined by the typical CCSN initial explosion energy and progenitor metallicity. For the optimistic cases in which the typical initial explosion energy is low, the diffuse flux of DNNB is comparable to the diffuse supernova neutrino background, which might be detected by the upcoming larger neutrino detectors such as Hyper-Kamiokande, JUNO, and DUNE. Moreover, the strong outflows from NDAFs could dramatically decrease their contribution to the neutrino background.
\end{abstract}


\section{Introduction}

At the end of their lives, massive stars ($ \gtrsim 8~{M}_{\odot}$) undergo core collapse. For a successful core-collapse supernova (CCSN) with progenitor mass $\lesssim 25~{M}_{\odot}$, core collapse gives birth to a hot proto-neutron star (PNS) and launches a shock wave that propagates through the stellar envelope and ejects heavy elements in the order of a few solar masses. Eventually, the PNS settles to a neutron star (NS). For a failed or a more massive CCSN, a black hole (BH) formation may occur due to the fallback process. Numerous CCSN simulations have confirmed the existence of fallback and studied its dynamics and effects \citep[see e.g.,][]{Bisnovatyi1984,Woosley1989,Chevalier1989,Woosley1995,Fryer1999,Fryer2009,MacFadyen2001,Zhang2008,Moriya2010,Moriya2019,Dexter2013,Wong2014,Perna2014,Branch2017,Chan2020}. The intensity of the fallback depends on the initial CCSN explosion energy and the binding energy of the star \citep[e.g.,][]{Fryer2006}. Strong fallback is expected for low initial explosion energies comparable to the binding energy of the star envelope. The fallback might be related to many observed phenomena: peculiar supernovae (SNe), long-duration gamma-ray bursts (LGRBs), late-time neutrino emission, and nucleosynthesis \citep[e.g.,][]{Wong2014,Liu2018,Song2019}.

In the collapsar model \citep[e.g.,][]{Woosley1993,MacFadyen1999} for LGRBs, a stellar-mass BH surrounded by a hyperaccretion disk forms by fallback, which is believed to launch the powerful jets, and the jets break out from the envelope to trigger an LGRB. Such a BH hyperaccretion system can launch relativistic jets via two mechanisms: the neutrino annihilation process and the Blandford-Znajek (BZ) mechanism \citep{Blandford1977}. If the mass accretion rate is very high ($\gtrsim$ 0.001 $M_{\odot} ~\rm s^{-1}$), the inner region of the disk is extremely hot and dense, and photons are trapped. The disk can only be cooled by neutrino emission. Neutrinos and antineutrinos are emitted from the disk surface and annihilate in the space out of the disk to produce relativistic jets. Such a disk is called neutrino-dominated accretion flow (NDAF), which has been widely studied in recent decades \citep[see e.g.,][]{Popham1999,Narayan2001,Kohri2002,Lee2005,Gu2006,Chen2007, Janiuk2007,Kawanaka2007,Liu2007,Liu2014,Lei2009,Xue2013,Song2016}, and for a review see \citet{Liu2017}.

MeV neutrino emission of NDAFs mainly depends on the mass accretion rates and the properties of central BHs \citep[e.g.,][]{Liu2016,Wei2019,Wei2021,Wei2022,Song2020,Qi2022}. It is generally considered that NDAFs widely exist in the center of massive collapsars and lasting tens to hundreds of seconds in the initial accretion phase \citep[e.g.,][]{Liu2018}. These neutrinos contain the information that can reveal the nature of central engines better than the multi-band radiation of gamma-ray bursts (GRBs). Many previous works have studied the detectability of neutrinos from NDAFs \citep[e.g.,][]{Nagataki2002,McLaughlin2007,Sekiguchi2011,Kotake2012,Caballero2012,Caballero2016,Liu2016}. If an NDAF event occurs in the Local Group, numerous neutrinos can be detected by the future facilities such as Hyper-kamiokande \citep[Hyper-K,][]{Abe2011}, Deep Underground Neutrino Experiment \citep[DUNE,][]{Abed2021}, and Jiangmen Underground Neutrino Observatory \citep[JUNO,][]{An2016}. However, the occurrence rate is a few per century \citep{Liu2016} and the only remedy is to wait. At even larger distances, the combined flux of neutrinos from all past NDAFs generates the diffuse NDAF neutrino background (DNNB). This neutrino background will contribute diffuse flux to the cosmic MeV neutrino background (CMNB), which is from all types of extragalactic MeV sources.

Until now, the most studied contribution to the CMNB is the diffuse SN neutrino background (DSNB). Over the years, the predictions of the DSNB have improved steadily \citep[see e.g.,][]{Horiuchi2009,Lunardini2009,Galais2010,Nakazato2013,Nakazato2015,Yuksel2015,Horiuchi2018,Muller2018,Abe2021,Kresse2021,Horiuchi2021,Li2021,Libanov2022,Tabrizi2021,Ashida2022,Ekanger2022,Ekanger2023,Gouvea2022,Anandagoda2023,Ashida2023}; and for the related reviews see \citet{Ando2004}, \citet{Beacom2010}, \citet{Lunardini2016}, \citet{Vitagliano2020}, and \citet{Suliga2022}. Compared to the DSNB, very few works focus on DNNB. \citet{Nagataki2003} first studied the neutrino background due to accretion disks in the frame of GRBs. They assumed that the rate of NDAFs traces the GRB formation history. They found that the predicted background neutrino flux might be detected by Totally Immersible Tank Assaying Nucleon Decay (TITAND) for the optimistic cases with high mass accretion rate. \citet{Schilbach2019} used updated models to study the CMNB from accretion disks formed during massive collapsars and compact object mergers. They adopted some typical mass accretion rates to calculate the neutrino emission of disks and found that DNNB is comparable (larger for high mass accretion rates) to DSNB. However, these two works did not consider the effects of progenitor properties and the initial explosion energy on the neutrino spectrum of NDAF. Recently, we have revealed how the neutrino emission of NDAFs depends on the initial explosion energies, masses, and metallicities of the progenitor stars in detail through a series of fallback CCSN simulations \citep[e.g.,][]{Liu2021,Wei2022}. In this paper, we aim to improve DNNB prediction by including those dependencies.

The paper is organized as follows. In Section 2, we describe the setup of our fallback CCSN simulations and discuss the effects of the initial explosion energies and the masses and metallicities of the progenitor stars on the neutrino emission of NDAFs. Section 3 is dedicated to our approach of formulating the DNNB. We discuss how the DNNB depends on the initial explosion energy and progenitor metallicity. Finally, we give the predicted neutrino event rates of DNNB at Super-K. In Section 4, we compared DNNB with DNSB. The conclusions and discussion are made in Section 5.

\section{CCSN simulations and NDAF emission}

\subsection{CCSN simulations}

Here we introduce the set-up of our fallback CCSN simulation. We adopt the pre-SN models as progenitor models \citep[e.g.,][]{Woosley2002,Woosley2007,Heger2010}. These models are nonrotating single stars, evolved using KEPLER code \citep{Weaver1978,Woosley2002} up to the onset of iron-core collapse. For these models, the ones with zero metallicity ($Z/Z_{\odot} = 0$) and solar metallicity ($Z/Z_{\odot} = 1$) are referenced from \citet{Heger2010} and \citet{Woosley2007}, respectively, as well as the ones with metallicity $Z/Z_{\odot} =0.1$ and $0.01$ are provided by Prof. Alexander Heger in private communication, where $Z_\odot$ represents the solar metallicity. Besides, we add pre-SN models with metallicity $Z/Z_{\odot} = 0.5$ as a supplement to investigate the behavior near the solar metallicity. These models are calculated by MESA code \citep{Paxton2011,Paxton2013}. We test that the pre-SN data from KEPLER and MESA show little difference, which has also been studied in  \citet{Sukhbold2014}. Masses included in this work were 20-40 $M_{\odot}$. Our stellar collapse and explosion simulations \citep{Liu2021,Wei2022} are performed in a series of 1D simulations with the Athena++ code \citep{White2016}.

Following \citet{Woosley1995} and \citet{Woosley2002}, we adopt the piston approach to carry out spherically symmetric explosion simulations. For each star, a piston was initially located at the outer edge of the iron core. At the beginning of the core collapse, the piston first moves inward for 0.45 s until it reaches a certain small radius, and then moves outward with a high velocity and decelerates smoothly until stopping at $10^{9}$ cm. The initial explosion conditions at the inner boundary are determined by the motion of the piston.

For each star, the simulation is divided into two steps to reflect the initial collapse and the subsequent explosion accompanied by the fallback process. In the first step, we mapped the structural profiles of progenitor stars into the Athena++ code. The computational domain has an inner boundary at $10^{9}$ cm, and an outer boundary is set at the surface of the progenitor star. A logarithmic grid with $10^{4}$ cells is used for the radial direction. A unidirectional outflowing inner boundary condition was used at the inner boundary to mimic the suction effect resulting from the hypothetical piston moving inwards. The simulation is run to 0.45 s, which reflects a brief free collapse of the star before the explosion. Then, the piston moves outward, which corresponds to the outward propagation of the blast. It takes, on average, approximately 1 s for a typical blast wave to reach $10^{9}$ cm \citep{Burrows2020}, which is also the period in which the piston moves outward. During this brief period, the gas flow would not change much. Therefore, we directly map the results of the first step to the new grid for the second step.

In the second step, the same outflowing inner boundary condition is set at $10^{9}$ cm. However, the outer boundary is set on a location far from the star surface ($\sim 10^{16}$ cm). At the beginning of the second step, the additional mass and energy are injected into the innermost cell adjacent to the inner boundary to mimic the outward blast passing through the inner boundary. The injected mass consists of two parts: one comes from the recording of inhaled mass during $0.45$ s collapse in the first step; another is the mass within the inner boundary ($\sim 10^{9}$ cm) excluding the mass of the iron core. The injected energy is just the setting explosion energy, which is assumed to have three values for each case, i.e., 2, 4, and 8$B$ ($1B=10^{51}~\rm erg$).

The grid for the second step has 2,000 logarithmic cells. For all cases, the simulation was run until the remnant growth ceased. In this step, the typical time-step amount for simulations is approximately 0.02 s, which is a very high time resolution for the physical process that occurs far from the central compact remnant. For further details of simulations, see \citet{Liu2021} and \citet{Wei2022}.

\subsection{BH evolution}

For each CCSN simulation, we obtain the evolution of the fallback mass supply rate. If the outflow is ignored, we can roughly consider the mass supply rate as the mass accretion rate. At the initial accretion stage, if the mass accretion rate is high ($\gtrsim 0.001 ~M_{\odot}~\rm s^{-1}$), the hyperaccretion disk would be in the state of NDAFs. If a BH is surrounded by a hyperaccretion disk, the mass and spin of this BH would be significantly time-dependent. The BH might be spun up by accretion and spun down by the BZ mechanism \citep[e.g.,][]{Lee2000a,Lee2000b,Wu2013,Qu2022,Li2024}. Since the BZ mechanism is not included in this work, according to the conversion of the energy and angular momentum, the evolution equations of a spinning BH can be expressed as \citep[e.g.,][]{Song2015}
\beq
\frac{dM_{\rm{BH}}}{dt}=\dot{M}e_{\rm{ms}},
\eeq
and
\beq
\frac{dJ_{\rm{BH}}}{dt}=\dot{M}l_{\rm{ms}},
\eeq
where $M_{\rm{BH}}$, $J_{\rm{BH}}$, and $\dot{M}$ are the mass and angular momentum of the BH and the mass accretion rate, respectively. $e_{\rm{ms}}$ and $l_{\rm{ms}}$ are the specific energy and angular momentum at the marginally stable orbit, which are given as \citep[e.g.,][]{Hou2014},
\beq
e_{\rm{ms}}=\frac{1}{\sqrt{3x_{\rm{ms}}}}\left ( 4-\frac{3a_{*}}{\sqrt{x_{\rm{ms}}}} \right ),
\eeq
and
\beq
l_{\rm{ms}}=\frac{2\sqrt{3}GM_{\rm{BH}}}{c}\left ( 1-\frac{2a_{*}}{3\sqrt{x_{\rm{ms}}}} \right ),
\eeq
where $c$ is the speed of light in vacuum, $a_{*}\equiv cJ_{\rm{BH}}/GM_{\rm{BH}}^{2}$ is the dimensionless spin parameter of the BH. $x_{\rm{ms}}=3+Z_{2}-\sqrt{(3-Z_{1})(3+Z_{1}+2Z_{2})}$ is the dimensionless radius of the marginally stable orbit \citep{Bardeen1972,Kato2008}, where $Z_{1}=1+(1-a_{*}^{2})^{1/3}[(1+a_{*})^{1/3}+(1-a_{*})^{1/3}]$, $Z_{2}=\sqrt{3a_{*}^{2}+Z_{1}^{2}}$ for $0< a_{*}< 1$.
Based on Equations (1)-(4), the evolution of the BH spin can be given by
\beq
\frac{da_{*}}{dt}=\frac{2\sqrt{3}\dot{M}}{M_{\rm{BH}}}\left ( 1- \frac{a_{*}}{\sqrt{x_{\rm{ms}}}}\right )^{2}.
\eeq
In this study, the initial BH spin parameter is set as $a_{*}=0.9$. The starting time is set at the time when the initial BH mass (core mass) reaches 2.3 $M_{\odot}$.

\subsection{NDAF emission}

\begin{figure}
\centering
\includegraphics[angle=0,scale=0.35]{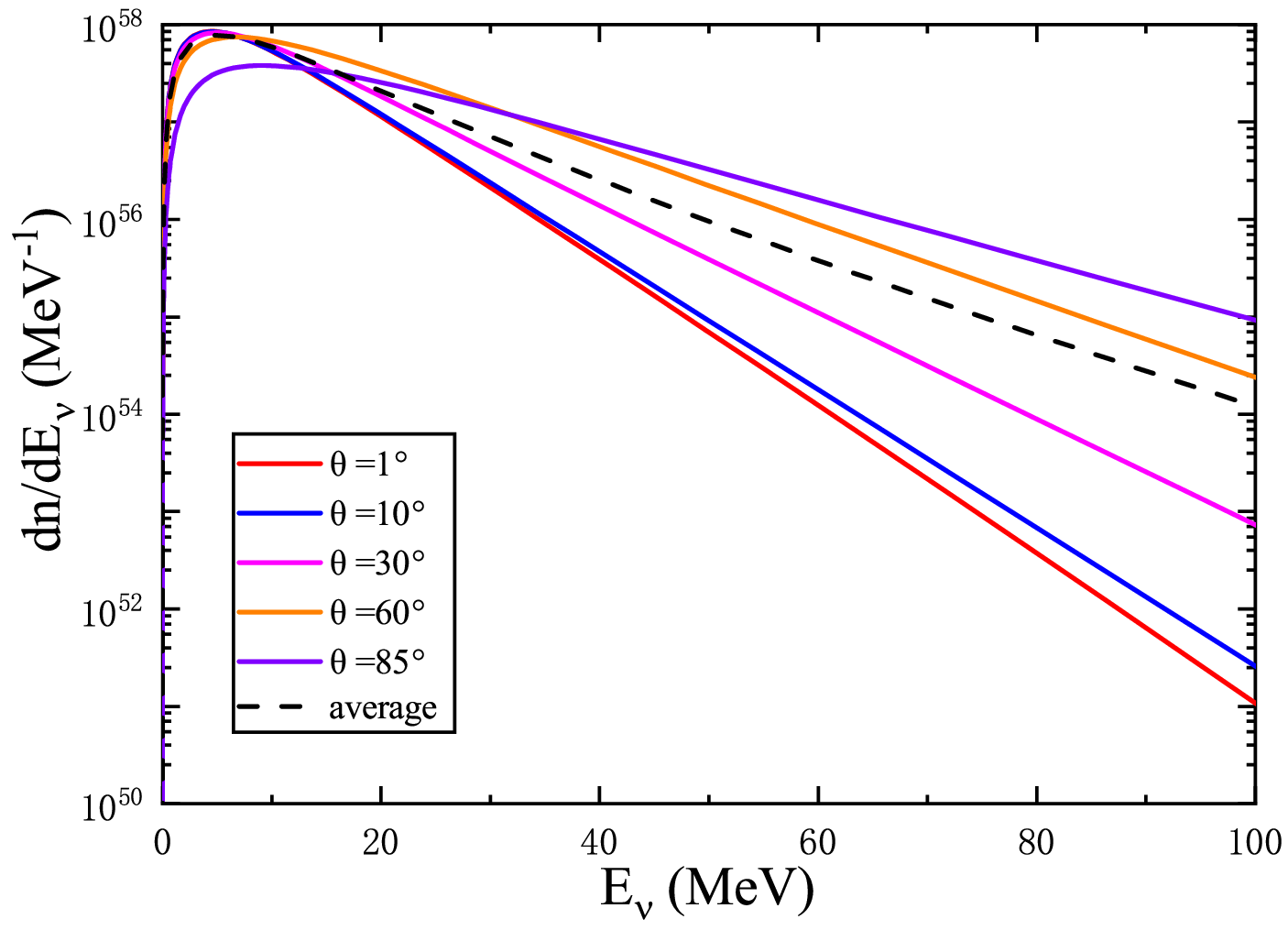}
\caption{Time-integrated electron antineutrino spectra of NDAFs for the different viewing angle $\theta$. The dashed line corresponds to the angle average neutrino spectrum. The progenitor mass and metallicity and the initial explosion energy are set as $40\;M_{\odot}$, $Z/Z_{\odot}=0.01$, and 2$B$, respectively.}
\end{figure}

The dominant neutrino-cooling process of NDAFs is the Urca process, and electron neutrinos and antineutrinos are the main products. In water Cherenkov detectors, the main detection channel is inverse beta decay (IBD) of electron antineutrinos:
\beq
\bar{\nu} _{e} +p\rightarrow e^{+} +  n.
\eeq
Thus, we focus on electron antineutrino emission of NDAFs throughout the paper.

According to the global solutions of NDAFs in \citet{Xue2013}, we derived the fitting formulae for the mean cooling rate due to electron antineutrino losses and the temperature of the disk as a function of the BH mass and spin, the mass accretion rate, and radius \citep{Wei2022},
\begin{align}
\log Q_{\bar{\nu}_{\rm{e}}}\;(\rm{erg}\;\rm{cm}^{-2}\;s^{-1}) =\ &41.17-0.21m_{\rm{BH}}+0.40a_{*}
\notag
\\&+1.69\log \dot{m}-3.96\log r,
\end{align}
and
\begin{align}
\log T \;(\rm{K})=\ &11.23-0.04m_{\rm{BH}}+0.10a_{*}
\notag
\\&+0.23\log \dot{m}-0.86\log r,
\end{align}
where $m_{\rm{BH}}=M_{\rm{BH}}/M_{\odot }$, $\dot{m}=\dot{M}/M_\odot~\rm s^{-1}$, and $r=R/R_g$ are the dimensionless BH mass, accretion rate, and radius, respectively. $R_g=2GM_{\rm{BH}}/c^2$ is the Schwarzschild radius.

For NDAFs, the neutrino-cooling rate in disk decreases with radius due to the drop of density and temperature. Therefore, neutrinos are mainly emitted from the inner region of the disk, and the observed spectra are affected by general relativistic effects. As a result, the observed neutrino spectrum is affected by the viewing angle. Here, the well-known ray-tracing method is used to calculate neutrino propagation effects. We treat the neutrino propagation in a manner similar to the photon propagation near the accreting BH. In this method, we divide the image of the accretion disk on the observer's sky into a number of small pixels. For each pixels, the position of the emitter on the accretion disk can be traced numerically based on the null geodesic equation \citep{Carter1968}. The neutrino emission by the disk at that point is calculated. Meanwhile, the energy shift of the neutrino emitted by the disk at that point can be calculated by taking into account the corresponding velocity and gravitational potential of this emission location. The observed neutrino flux contributed by each pixel is thus obtained, and summing over all the pixels gives the total observed flux distribution as
\beq
f_{{E}_{\rm{obs}}}=\int_{\rm image}g^{3}I_{E_{\rm{em}}}d\Omega _{\rm{obs}},
\eeq
where $E_{\rm{obs}}$ is the observed neutrino energy, $E_{\rm{em}}$ is the neutrino emission energy from the local disk, $g \equiv E_{\rm obs} / E_{\rm em}$ is the energy shift factor, and $\Omega _{\rm{obs}}$ is the solid angle of the disk image to the observer. $I_{E_{\rm{em}}}$ is the local emissivity, which can be calculated by the cooling rate $Q_{\bar{\nu}_{\rm{e}}}$ as
\beq
I_{E_{\rm{em}}}=Q_{\bar{\nu}_{\rm{e}}}\frac{f_{E_{\rm{em}}}}{\int f_{E_{\rm{em}}}dE_{\rm{em}}},
\eeq
where $f_{E_{\rm{em}}}=E_{\rm{em}}^{2}/[\exp (E_{\rm{em}}/kT-\eta )+1]$ is the unnormalized Fermi-Dirac spectrum \citep[e.g.,][]{Rauch1994,Fanton1997,Li2005,Liu2016}.

As an example, Figure 1 shows the effect of viewing angle on electron antineutrino spectra of NDAFs. Here, the mass and metallicity of the progenitor and the initial explosion energy are set as $40\;M_{\odot}$, $Z/Z_{\odot}=0.01$, and 2$B$, respectively. $\theta=90^{\circ}$ corresponds to the case that the observer is located in the equatorial plane. The viewing angle has significant effects on the high-energy range of spectra. This is because high-energy neutrinos are mainly produced in the inner region of the disk, and these neutrinos would be more affected by the general relativistic effects. As the viewing angle increases, the luminosity at high energy increases. For our DNNB flux predictions, we calculate the angle average neutrino spectrum of NDAF to fully take into account the viewing angle effects. The angle average neutrino spectrum per NDAF can be derived as
\beq
F(E_{\nu})=\frac{dn}{dE_{\nu}}=\sum_{i}\frac{\int_{\bigtriangleup \theta _{i}} d\theta}{\int_{0}^{\pi/2} d\theta} f_{i} (E_{\nu}),
\eeq
where $E_{\nu }$ is the energy of neutrino, $\bigtriangleup\theta_{i}$ is the angle range of angle bin $i$, and $f_{i} (E_{\nu })$ is the observed spectrum at the corresponding viewing angle. As shown in Figure 1, the dashed line represents the angle-averaged spectrum. In the subsequent calculations, we all adopt angle-averaged neutrino spectrum.

The effects of the progenitor metallicity on the neutrino spectra of NDAFs are displayed in Figure 2. Panels (a), (b), and (c) correspond to the cases of progenitors with masses of 20, 30, and 40 $M_{\odot}$, respectively. The initial explosion energy is 2$B$. The black, red, blue, orange, and pink curves correspond to the progenitor metallicities of $Z/Z_{\odot}=0$, 0.01, 0.1, 0.5, and 1, respectively. In Figure 2 (a), we only show neutrino spectra of NDAFs from progenitors with metallicity of $Z/Z_{\odot}=0.01$ and 0.1. This is because the final remnant of progenitors with metallicities of $Z/Z_{\odot}=0, 0.5$, and 1 are NSs rather than BHs in our simulations. Most of high-energy neutrinos are produced in the early hyperaccretion stage with a high accretion rate \citep[e.g.,][]{Wei2019,Wei2021}. For the same initial explosion energy, the mass accretion rate at the early stage is determined by the compactness of pre-SN core. The mass accretion rate at the early stage increases as the core becomes denser. The dependence of progenitor masses and metallicities on the structural characteristics of pre-SN stars have been studied by some previous works \citep[e.g.,][]{OConnor2011,Sukhbold2014}. These studies find a non-monotonic behavior for the compactness as a function of progenitor mass and metallicity. There are a lot of statistical variations over the mass range, more than as a function of metallicity. For a given core mass of a progenitor star, the evolution of the core is mostly independent of metallicity. Metallicity mainly affects the mass loss of massive stars and plays an important role in the envelope mass of stars. As a result, for different progenitor masses, the neutrino emission of NDAFs is not strictly dependent on metallicity. Besides, based on previous stellar evolutionary studies \citep{Sukhbold2014}, the core compactness parameters of solar metallicity stars are commonly smaller than those of low metallicity stars. Hence, solar metallicity is unfavorable for high-energy neutrino emission of NDAFs.

\begin{figure}
\centering
\includegraphics[angle=0,scale=0.35]{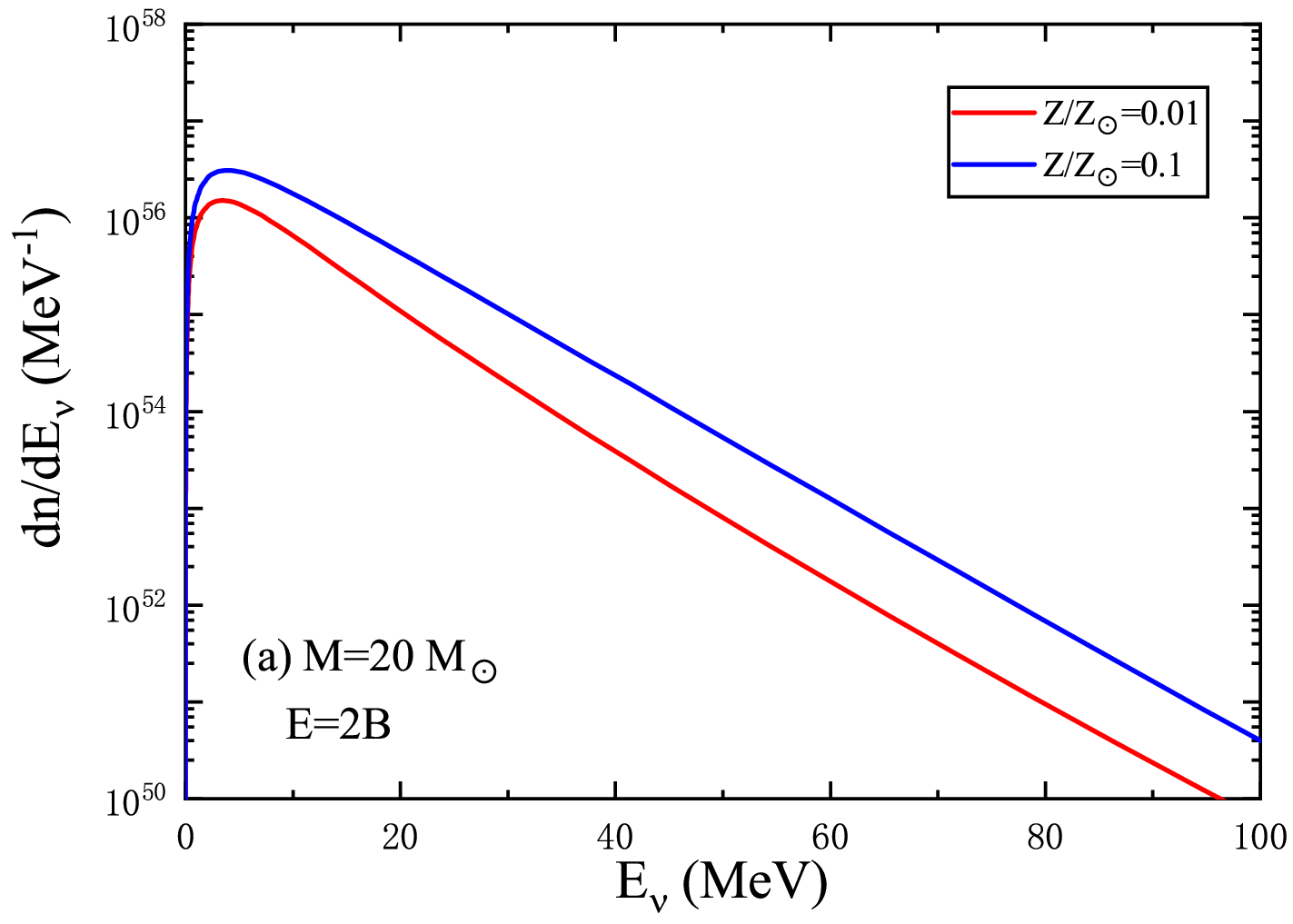}
\includegraphics[angle=0,scale=0.35]{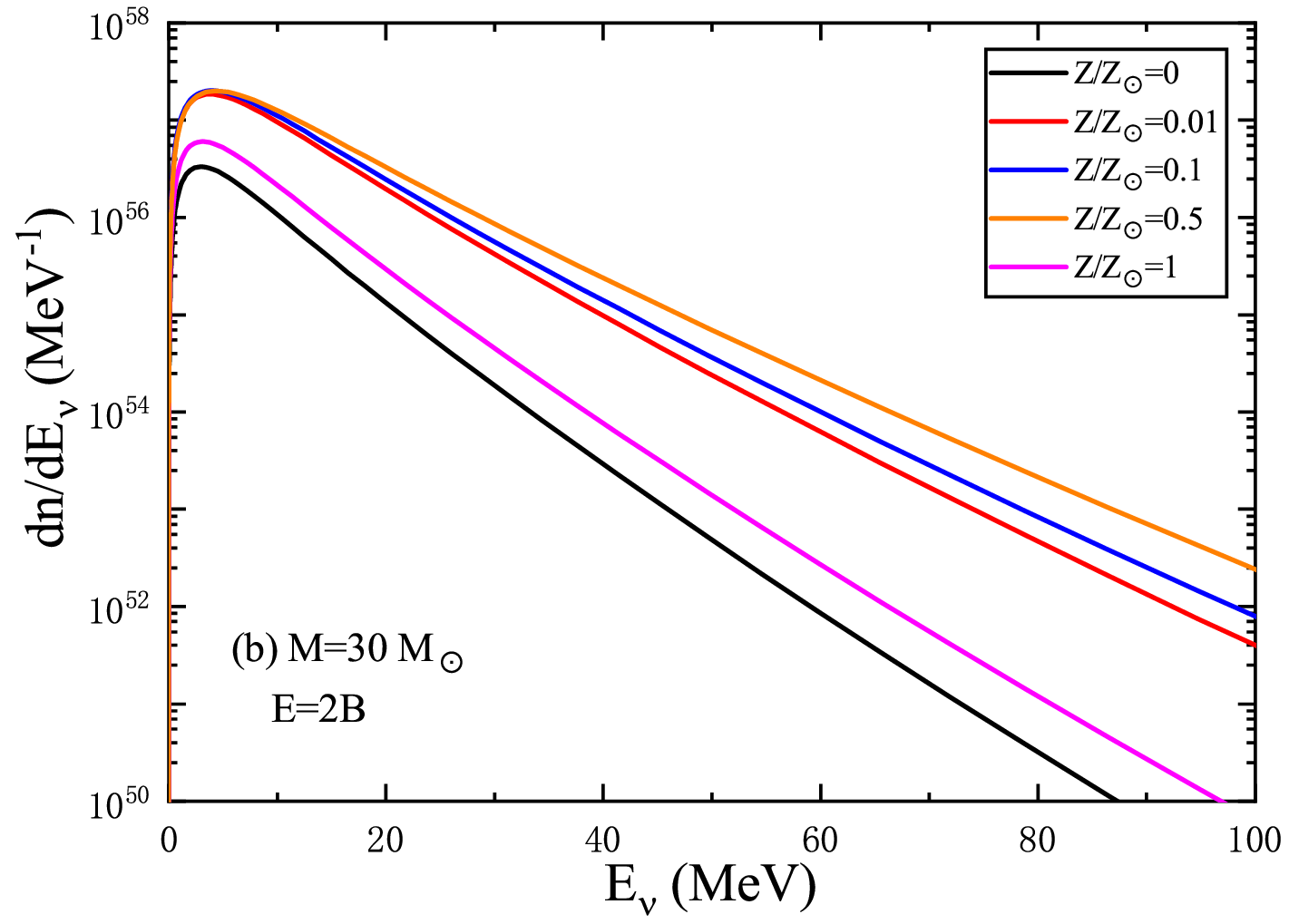}
\includegraphics[angle=0,scale=0.35]{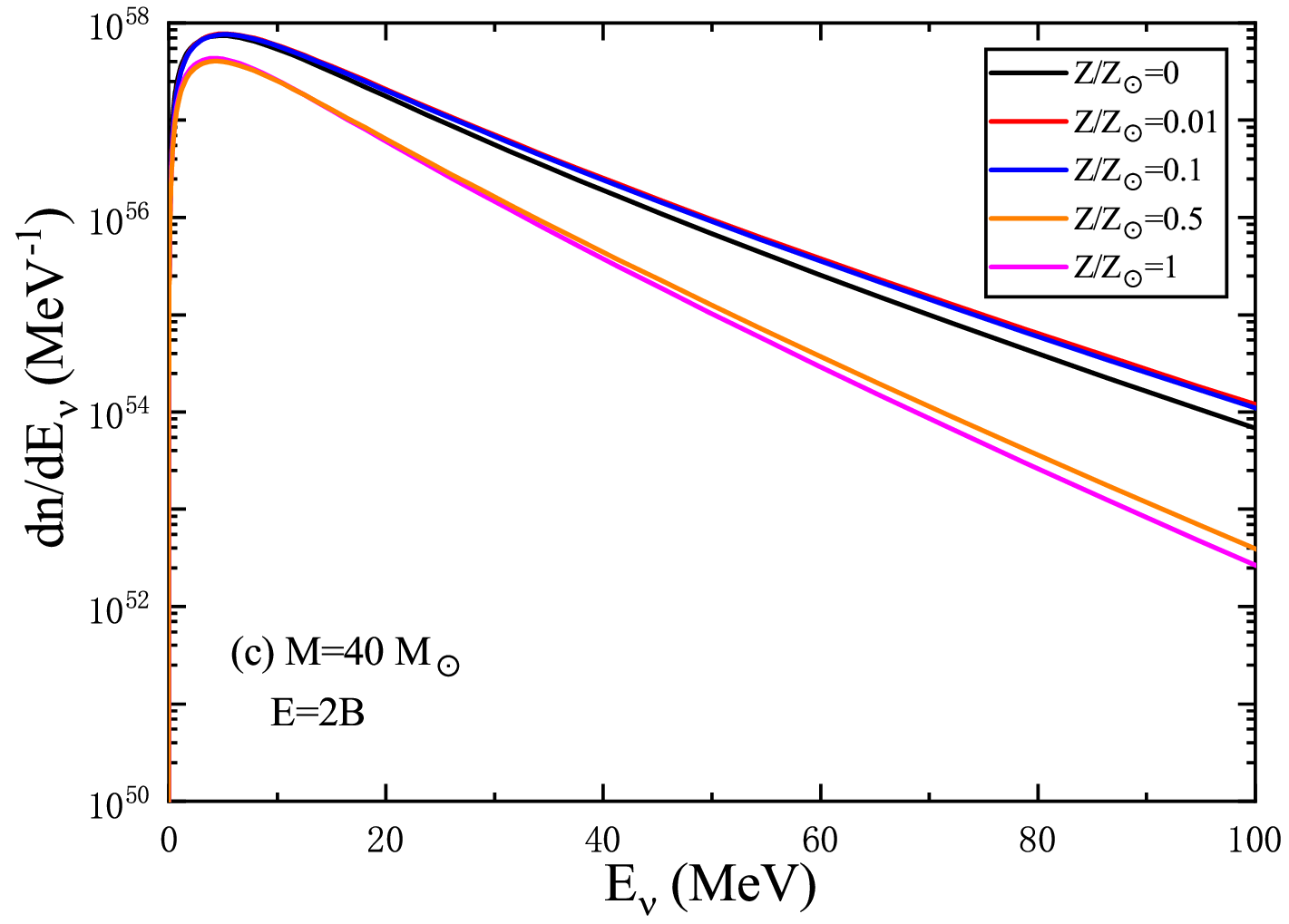}
\caption{Time-integrated electron antineutrino spectra of NDAFs with different masses and metallicites of progenitors. The black, red, blue, orange, and pink curves correspond to progenitor star metallicities of $Z/Z_{\odot}=0$, 0.01, 0.1, 0.5, and 1, respectively. The initial explosion energy is 2$B$.}
\end{figure}

Figure 3 shows the effects of initial explosion energy on neutrino spectra of NDAFs. The metallicity is set to $Z/Z_{\odot}=0.01$. The black, red, and blue curves correspond to the initial explosion energy of 2, 4, and 8$B$, respectively. As the explosion energy decreases, the amplitudes of the spectral lines increase. This is because the weaker explosion energy corresponds to a more powerful fallback. In Figures 3 (a) and (b), the case of 8$B$ is not shown because the final remnants of these progenitors are both NSs.

\begin{figure}
\centering
\includegraphics[angle=0,scale=0.35]{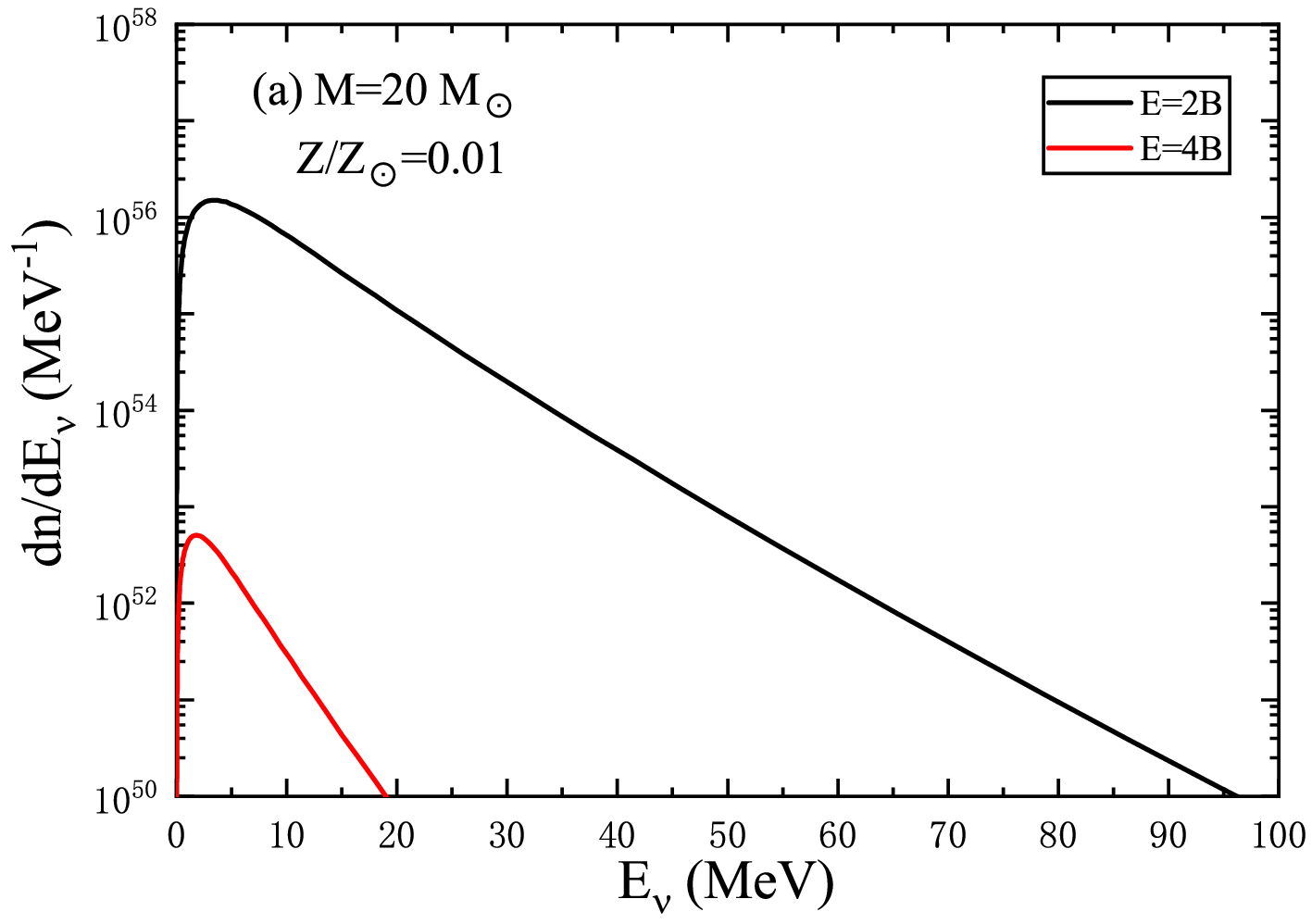}
\includegraphics[angle=0,scale=0.35]{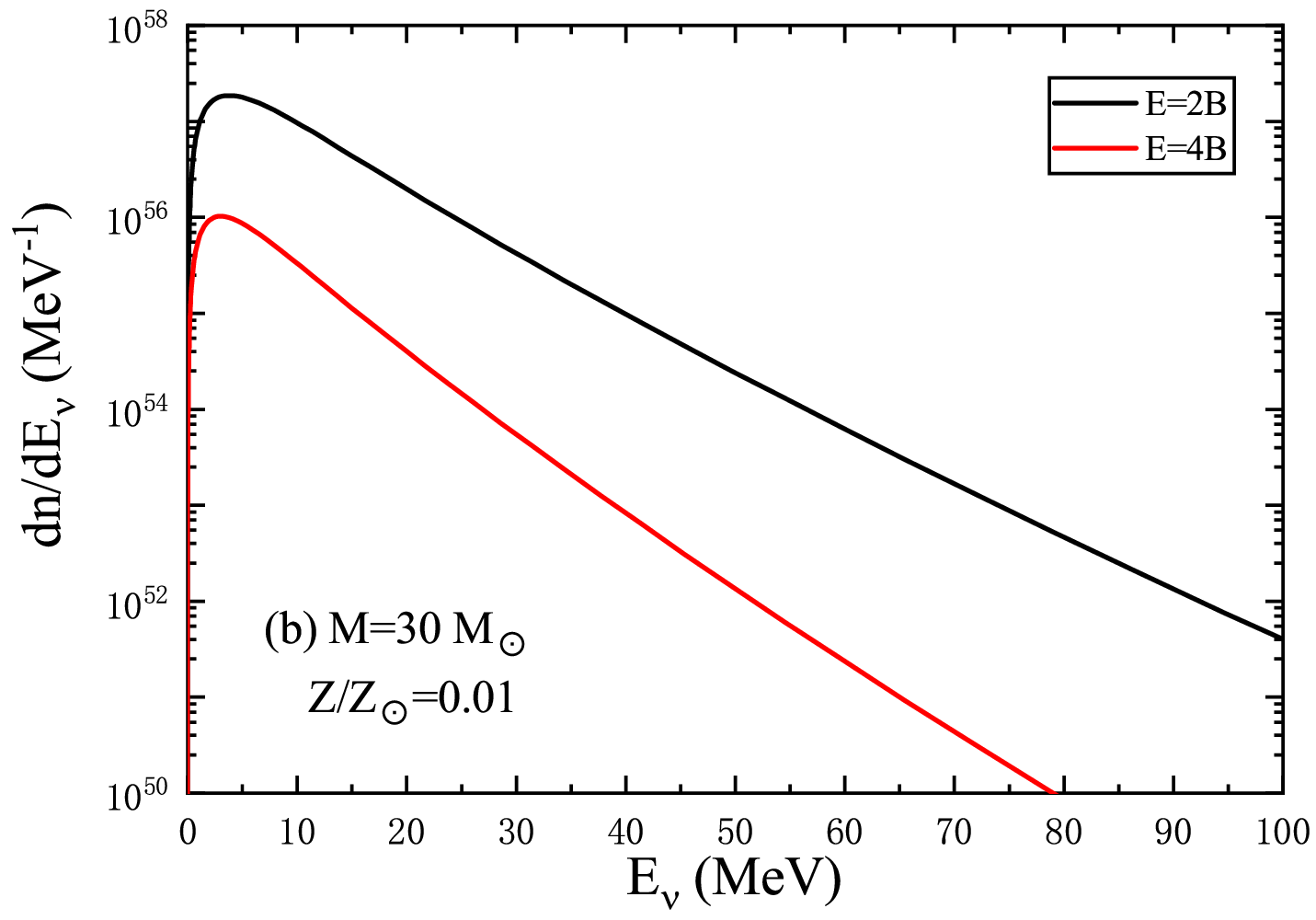}
\includegraphics[angle=0,scale=0.35]{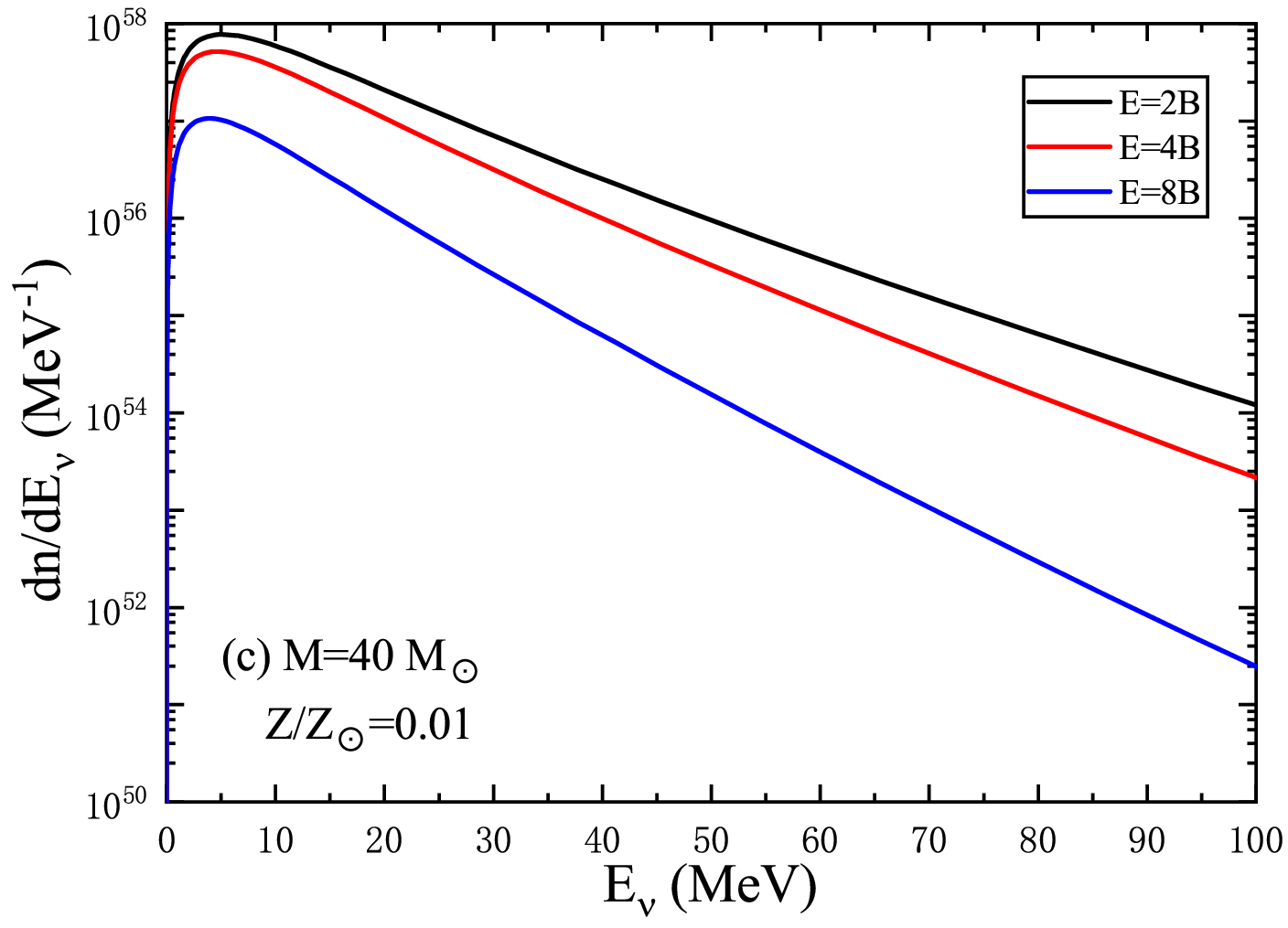}
\caption{Time-integrated electron antineutrino spectra of NDAFs with different initial explosion energy. The black, red, and bule curves correspond to initial explosion energies of 2, 4, and 8$B$, respectively. The metallicity is set as $Z/Z_{\odot}=0.01$.}
\end{figure}

\section{DNNB prediction and detection}

A prediction of the DNNB requires a good understanding of the average neutrino emission spectrum and the occurrence rate of NDAFs. As shown in the previous section, the properties of progenitors and the initial explosion energy would determine whether the fallback process produces NDAFs. Here, we investigate the effects of metallicities and initial explosion energies of progenitors on DNNB.

\begin{table}
	\caption{$M_{\rm{min}}$ for different metallicities and initial explosion energies.}
	\label{table1}
	\begin{tabular}{ccc}
		\hline
		Metallicity   &  Initial Explosion Energy & $M_{\rm{min}}$    \\
        ($Z/Z_\odot$)   & ($B$)     & ($M_\odot$)                 \\
        \hline
		 0	&	2	&	30	\\
         0.01	&	2	&	20	\\
         0.1	&	2	&	20	\\
         0.5	&	2	&	30	\\
         1	&	2	&	30	\\
         0.01	&	4	&	20	\\
         0.01	&	8	&	40	\\
		\hline
	\end{tabular}
\end{table}

\subsection{Cosmic NDAF history}

The progenitors of CCSNe have relatively short lifetimes ($\lesssim 10^{8}$ years) compared to cosmic timescales \citep{Kennicutt1998}. As a result, the rate of NDAFs can be calculated by using the star formation rate and initial mass function (IMF). The cosmic NDAF rate at a redshift of $z$ is calculated as
\beq
R_{\rm{NDAF}} (z)=R_{\rm{SFR}}(z)\frac{\int_{M_{\rm{min}}}^{M_{\rm{max}}} \Psi (M)dM}{\int_{0.1}^{125} M\Psi (M)dM},
\eeq
where $R_{\rm{SFR}}(z)$ is the cosmic star formation rate in units of $\rm{Mpc}^{-3}\; year^{-1}$, which can be deduced from observations \citep[e.g.,][]{Hopkins2006,Reddy2008,Rujopakarn2010}. Here, we adopt the continuous broken power law description by \citet{Yuksel2008},
\beq
R_{\rm{SFR}} (z)=\dot{\rho} _{0}\left [ (1+z)^{\alpha\eta } +(\frac{1+z}{C} )^{\beta \eta } + (\frac{1+z}{D} )^{\gamma \eta }\right ]^{1/\eta },
\eeq
where $\alpha =3.4$, $\beta =-0.3$, $\gamma =-2$, $\eta =-10$, $C\simeq 5100$, $D\simeq 14$, and $\dot{\rho} _{0}=0.014 ~ \rm{Mpc}^{-3}~ year^{-1}$. In this work, we use the Salpeter initial mass function \citep{Salpeter1955}, $\Psi (M)\propto M^{-2.35}$, with a mass range of $0.1-125~M_{\odot}$. Here, $M_{\rm{max}}$ and $M_{\rm{min}}$ are the maximum and minimum masses of progenitors that produce NDAFs, respectively. Due to the influence of the metallicity of the progenitor and initial explosion energy on the fallback process of CCSNe, not all CCSNe can produce NDAFs. Especially, $M_{\rm{min}}$ depends on the metallicity of progenitor and initial explosion energy \citep{Liu2021}. Based on the results of our simulations, we gave $M_{\rm{min}}$ for different metallicity and initial explosion energy, and the results are listed in Table \ref{table1}. The DNNB prediction depends weakly on the $M_{\rm{max}}$ and we set $M_{\rm{max}}=50\;M_{\odot}$, which is the upper limit of the progenitor mass for NDAFs \citep{Liu2021b}.

\subsection{IMF-weighted average neutrino spectrum}

The average neutrino emission spectrum for a population of progenitors is computed by weighting each progenitor by
\beq
\frac{dN}{dE_{\nu}} =\sum_{i}\frac{\int_{\bigtriangleup M _{i}} \Psi (M)dM}{\int_{M_{\rm{min}}}^{M_{\rm{max}}} \Psi (M)dM} F_{i} (E_{\nu }),
\eeq
where $\Psi (M)$ is once again the IMF, $\bigtriangleup M _{i}$ is the mass range of mass bin $i$, and $F_{i} (E_{\nu })$ is the neutrino spectrum of the NDAF with progenitor mass of $M_{i}$.

\subsection{DNNB flux}

The DNNB flux spectrum at the Earth is calculated from the IMF-weighted neutrino spectrum of past NDAF $dN/dE_{\nu}$ and the evolving NDAF rate $R_{\rm{NDAF}} (z)$ as
\beq
\frac{d\Phi }{dE_{\nu}} =c\int_{0}^{\infty } (1+z)\frac{dN}{dE_{\nu} ^{'}}R_{\rm{NDAF}}(z)\left |\frac{dt}{dz} \right |dz,
\eeq
where $E_{\nu}^{'} =E_{\nu}(1+z)$, and $\left |dz/dt \right |=H_{0}(1+z)\sqrt{\Omega_{\Lambda}+\Omega _{m}(1+z)^{3}}$. Here, $H_{0}= 70\;\rm{km} \;\rm{s}^{-1}$, $\Omega _{m}=0.3$, and $\Omega_{\Lambda}=0.7$ are adopted. We set $z_{\rm{max}}=5$, which is large enough to incorporate the majority of the DNNB flux.

\begin{figure}
\centering
\includegraphics[angle=0,scale=0.35]{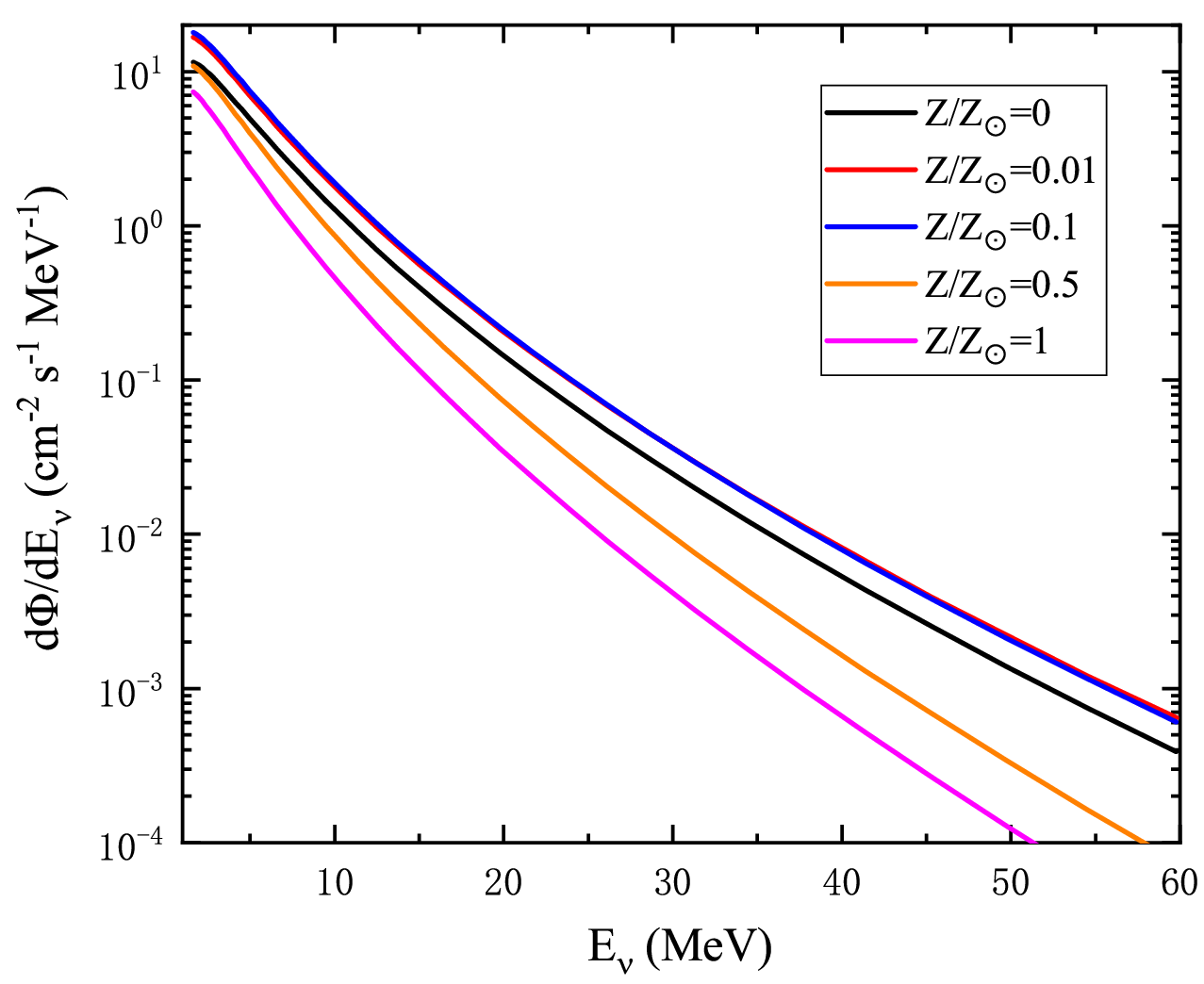}
\caption{$\bar{\nu} _{e}$ flux of DNNB as a function of the neutrino energy for different metallicities with the initial explosion energy 2$B$.}
\end{figure}

Figure 4 shows the effect of the metallicity of the progenitor on the DNNB flux spectrum. The initial explosion energy of all progenitors is set as 2$B$. The black, red, blue, orange, and pink curves correspond to $Z/Z_{\odot}=0$, $0.01$, $0.1$, $0.5$, and $1$, respectively. The amplitudes of the spectra lines are not monotonically dependent on metallicity. Solar metallicity is not beneficial for the detection of DNNB. The DNNB flux spectrum is determined by the IMF-weighted neutrino spectrum of past NDAF and the cosmic rate of NDAF. As illustrated in Figure 2, solar metallicity is unfavorable for the emission of high-energy neutrinos for progenitors with different masses. Besides, as listed in Table 1, a progenitor with solar metallicity has a large $M_{\rm{min}}$.

In Figure 5, we display the effects of the initial explosion energy on the DNNB flux spectra. We assume that all progenitors have the same metallicity ($Z/Z_{\odot}=0.01$). The black, red, and blue curves correspond to the initial explosion energy of 2, 4, and 8$B$, respectively. The weaker explosion energy corresponds to a more powerful fallback, which enhances the neutrino emission and the event rate of NDAFs.

\begin{figure}
\centering
\includegraphics[angle=0,scale=0.35]{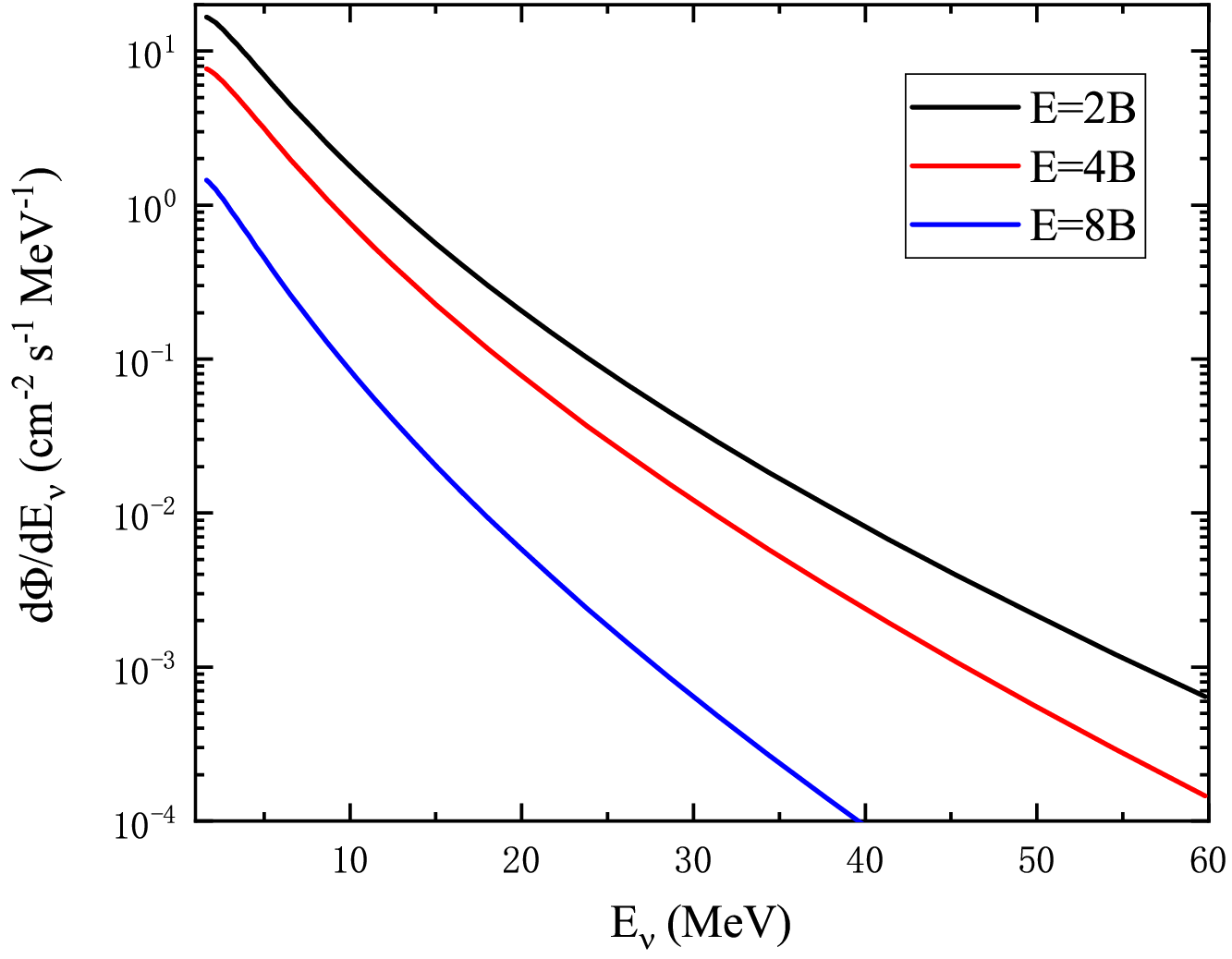}
\caption{$\bar{\nu} _{e}$ flux of DNNB as a function of the neutrino energy for different initial explosion energies with $Z/Z_{\odot}=0.01$.}
\end{figure}

\subsection{Detection rates}

We calculate the event rates of DNNB at Super-K per year as
\beq
\frac{dN_{\rm e^+} }{dE_{\rm e^+} }(E_{\rm e^+}) =N_{\rm{t}} \sigma (E_{\bar{\nu}_{e}}) \frac{d\Phi }{dE_{\bar{\nu}_{e}}},
\eeq
where $\sigma (E_{\bar{\nu}_{e}})$ is the IBD cross section \citep{Vogel1999,Strumia2003}. The positron energy is given as $E_{\rm e^+}= E_{\bar{\nu}_{e}} -\Delta c^{2}$, where $\Delta$ is the neutron-proton mass difference. $N_{\rm{t}}$ is the number of free protons contained in the detector. For Super-K (22.5 kton), $N_{\rm{t}}=1.5\times 10^{33}$ \citep{Horiuchi2009}.

\begin{figure*}
\centering
\includegraphics[angle=0,scale=0.35]{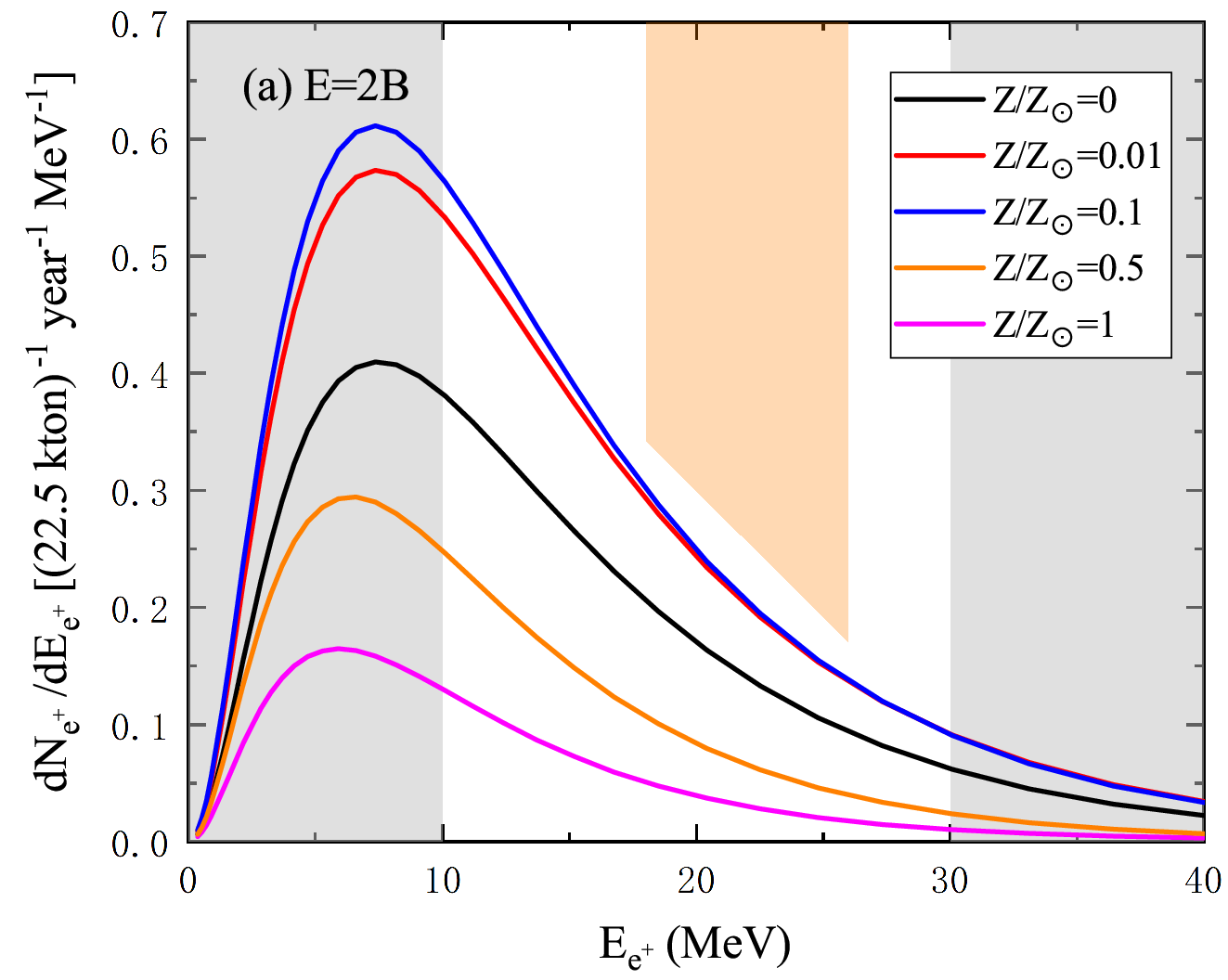}
\includegraphics[angle=0,scale=0.35]{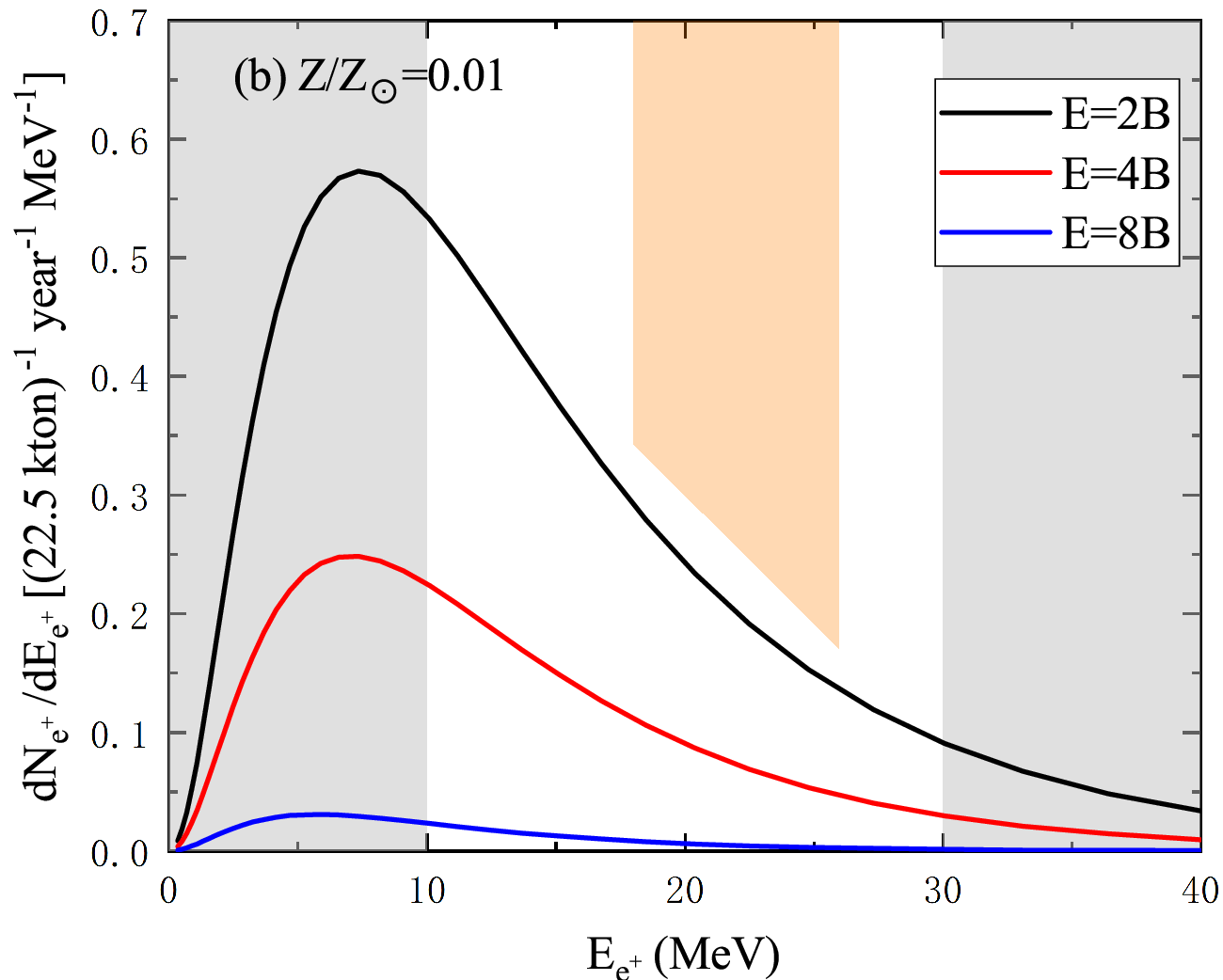}
\caption{DNNB event rates in 22.5 kton Super-K (flux spectra weighted with the detection cross section) against positron energy. Left and right panels show the effects of progenitor metallicities and initial explosion energies on event rate spectra, respectively. The gray regions correspond to those where the expected number of background events is dominant. The inferred limits based on 2003 Super-K data are indicated by orange shaded region.}
\end{figure*}

The positron spectra evaluated for Super-K per year obtained from our models with different progenitor metallicity and initial explosion energy are shown in Figure 6. In both panels, the gray shadings bracket the energy window ($\sim$ 10-30 MeV), which is relevant for DNNB detection in future neutrino observations. The shadings indicate backgrounds (such as atmospheric neutrinos at high energies and reactor and solar neutrinos at low energies), which dominate the flux and make DNNB detection unfeasible \citep{Lunardini2016,Kresse2021}. The orange-shaded region denotes the 2003 upper limit by Super-K, $<2$ events $(22.5\;\rm{kton}\; \rm{year})^{-1} $ in the energy range of 18-26 MeV \citep{Horiuchi2009}. The effects of progenitor metallicity and initial explosion energy on the DNNB detection spectrum follow those of the factors on the DNNB flux. The initial explosion energy is crucial for detection. If CCSNe with low initial explosion energies are universal, DNNB is hopefully to be detected by upcoming neutrino detectors, such as Hyper-K, JUNO, and DUNE. Note that if solar metallicity is universal, the measurement of DNNB would be difficult.

\section{DNNB v.s. DSNB}

Note that there are still many uncertainties in the current prediction of DSNB. For predictions from different groups, the DSNB flux can vary over one order of magnitude \citep{Abe2021}. Here, we mainly refer to the calculations in \citet{Lunardini2016}. The neutrino emission from a CCSN can be parameterized by the well-known $\alpha$-fit spectra \citep{Keil2003}, i.e.,
\beq
F_{\rm{CCSN}}(E_{\nu } )\simeq  \frac{(1+\alpha)^{1+\alpha} E^{\rm{tot}} _{\nu }}{\Gamma(1+\alpha) E_{0\nu } ^{2}} \left (\frac{E_{\nu } }{E_{0\nu }} \right )^{\alpha} e^{-(1+\alpha)E_{\nu } /E_{0\nu }},
\eeq
where $\Gamma(x)$ stands for the Gamma function, $E^{\rm{tot}} _{\nu }$ is the total energy of neutrinos emitted from CCSNe, $E_{0\nu}$ is the mean neutrino energy, and $\alpha$ is a a spectral shape parameter. Here, we use the typical values \citep{Keil2003}: $E_{0\nu}=15 \;\rm{MeV}$, $E^{\rm{tot}} _{\nu }=5\times 10^{52} \;\rm{erg}$, and $\alpha=3.5$.

Assuming that stars are distributed in mass according to the Salpeter initial mass function, the cosmic history of the comoving CCSN rate, $R_{\rm{CCSN}}(z)$, can be calculated as \citep{Beacom2010}
\beq
R_{\rm{CCSN}}(z)=R_{\rm{SFR}}(z)\frac{\int_{8}^{50} \Psi (M)dM}{\int_{0.1}^{100} M\Psi (M)dM}.
\eeq
Finally, the event rate spectrum in Super-K is obtained as
\beq
\begin{split}
\frac{dN_{\rm{e}^{+}} }{dE_{\rm{e}^{+}}} (E_{\rm{e}^{+}}) &=c N_{\rm{t}} \sigma (E_{\bar{\nu }_{e}}) \\
                     & \times \int_{0}^{\infty } (1+z)F_{\rm{CCSN}}(E_{\bar{\nu }_{e}}^{'})R_{\rm{CCSN}}(z)\left |\frac{dt}{dz}  \right |dz,
\end{split}
\eeq
where $E_{\bar{\nu }_{e}}^{'}=E_{\bar{\nu }_{e}}(1+z)$.

In Figure 7, we show the predicted positron spectrum of DSNB. As a comparison, the event rate spectra of DNNB corresponding to the initial explosion energy of 2, 4, and 8$B$ are also displayed in the figure. The typical metallicity is set to $Z/Z_{\odot}=0.01$. As discussed above, DNNB flux is mainly determined by the initial explosion energy. If weak explosion is universal, the flux of DNNB may exceed that of DSNB, and DNNB may dominate in CMNB. In some recent studies on DSNB, the contribution of diffuse flux from failed CCSNe (direct collapse into a BH without explosion) to DSNB has been widely investigated \citep[see e.g.,][]{Lunardini2009,Lunardini2016,Yuksel2015,Ashida2022,Nakazato2013,Kresse2021,Horiuchi2018}. Many studies have suggested that the neutrino emission of a failed CCSN is somewhat more luminous and decidedly more energetic than that of an NS-forming CCSN due to the rapid contraction of the newly formed PNS preceding the BH formation \citep[e.g.,][]{Sumiyoshi2006,Sumiyoshi2007,Sumiyoshi2008,Nakazato2008,Nakazato2013,Fischer2009}. The fraction of core collapses that result in failed SNe remains uncertain. In general, it will depend on metallicity, the equation of state, and the explosion mechanism. If the failed CCSN fraction is large, diffuse fluxes from failed CCSNe may significantly increase the flux of DSNB, especially in the high-energy range of the spectra. Moreover, the impact of late-time neutrino emission on the DSNB has been studied by some works \citep{Horiuchi2018,Kresse2021,Ekanger2022}. In CCSNe, the cooling of the PNS ($\gtrsim 1~$s after core bounce) is an important sources of neutrinos. \citet{Ekanger2022} found that the predicted DSNB event rate at Super-K can vary by a factor of $\sim$ 2-3 depending on the cooling-phase treatment. Their study suggests that improving the understanding of the late-phase neutrino emission will be crucial for the uncertainties in DSNB. Despite uncertainties, the contribution of DNNB to CMNB should be noticed, which would be of great service to the detection of CMNB.

\section{Conclusions and Discussion}

We have studied the spectra and the detection rates of DNNB based on fallback CCSN simulations. The effects of progenitor properties and initial explosion energies on DNNB are investigated. We found that they can affect the neutrino spectra and event rates of NDAFs. The neutrino emission of NDAFs depends on the fallback process in CCSNe. Lower initial explosion energy is beneficial for producing powerful fallback, resulting in stronger neutrino emission of NDAFs. Meanwhile, lower initial explosion energy may correspond to lower minimum masses of progenitors that produce NDAFs. Therefore, low initial explosion energy may enhance the event rates of NDAFs. The influence of metallicity on DNNB is not monotonic, and solar metallicity is not beneficial for the detection of DNNB.  Moreover, for weak initial explosion, the flux of DNNB is comparable (larger for low initial explosion energy) to DSNB one and provides a significant contribution to CMNB.

\begin{figure}
\centering
\includegraphics[angle=0,scale=0.35]{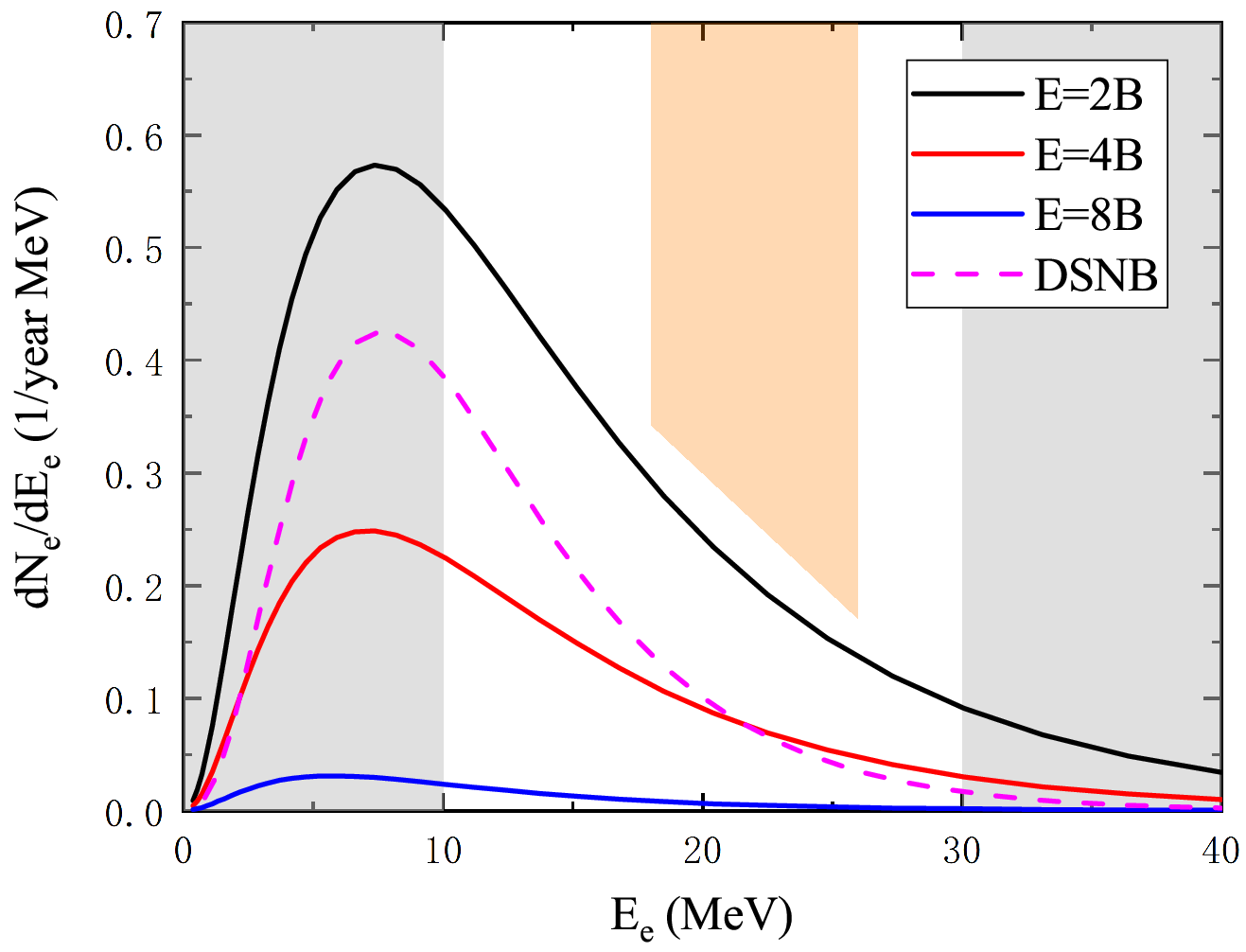}
\caption{Comparison of DNNB and DSNB. The black, red, and blue curves correspond to DNNB with initial explosion energies of 2, 4, and 8$B$, respectively. The metallicity is set to $Z/Z_{\odot}=0.01$. The pink dashed line corresponds to DSNB.}
\end{figure}

In the above framework, the predicted neutrino flux of DNNB varies as the values of the typical initial explosion energies and metallicities are varied. Cosmic metallicity evolution has been proven by the observations of galaxies \citep{Maiolino2008}. In general, the metallicity of galaxies will be lower at higher redshift universe. Numerous metal-free and very metal-poor massive stars are formed in the high redshift universe, and the frequency of solar metallicity stars in galaxies would be low in the early universe. The metal-rich stars might form in the central supermassive BH accretion disk and their evolution should contribute to the metallicity of active galactic nuclei \citep[e.g.,][]{Qi2022b}. In the low redshift universe, the metallicity of galaxies would be relatively high, but galaxies whose metallicities are about one order of magnitude lower than the solar value reside within the redshift $z<1$ \citep{Peeples2013}. This is beneficial for the detection of DNNB. The predictions of DNNB is highly sensitive to the typical initial explosion energy of CCSNe, which remains uncertain. In \citet{Liu2021}, we studied the possibility of the formation of the lower mass gap in compact object distribution by considering the fallback mechanism in the CCSN scenario. We found the shape of the gap is mainly determined by the typical initial explosion energy of CCSNe. The future multimessenger observations on compact objects in the lower mass gap may give a limitation on the typical initial explosive energy of CCSNe.

In our work, we have not taken into account diffuse flux from NDAFs produced in mergers. The neutrino emission of NDAFs in the scenario of NS-NS or NS-BH mergers can be comparable to that of NDAFs in collapsar scenario \citep{Qi2022}. However, cosmological merger rate are far lower than the event rate of collapsar. Hence, the diffuse flux from NDAFs produced in mergers is at least two orders of magnitude smaller than that from NDAFs produced in collapsars \citep{Schilbach2019}.

Moreover, we did not consider the outflows in our calculations. The disk outflows play an important role in critical accretion system \citep[e.g.][]{Liu2008,Liu2014,Gu2015}. \citet{Liu2021b} found that the neutrino luminosity of NDAFs with strong outflows would be at least one order of magnitude lower than that of NDAFs without outflows. Therefore, if powerful disk outflows are universal in NDAFs, the contribution of DNNB to CMNB would be reduced.

Future megaton-scale neutrino detectors like Hyper-K will open the door to the high-statistics CMNB detections. While the prediction of DNNB is limited by uncertainties, the contribution of DNNB to CMNB is noteworthy. The detection of DNNB would significantly improve the investigations of CCSNe and NDAFs.

\acknowledgments
We thank Prof. Alexander Heger for providing us pre-SN data and helpful discussion, and the anonymous referee for very useful suggestions and comments. This work was supported by the National Key R\&D Program of China (Grant No. 2023YFA1607902), the National Natural Science Foundation of China (Grant Nos. 12173031, 12221003, and 12303049), and the National Fund for Postdoctoral Researcher Program (Grant No. GZC20231424), the China Postdoctoral Science Foundation 2023M741998.


\begin{thebibliography}{99}
\bibitem[Abe et al.(2011)]{Abe2011} Abe, K., Abe, T., Aihara, H., et al.\ 2011, arXiv:1109.3262. doi:10.48550/arXiv.1109.3262
\bibitem[Abe et al.(2021)]{Abe2021} Abe, K., Bronner, C., Hayato, Y., et al.\ 2021, \prd, 104, 122002. doi:10.1103/PhysRevD.104.122002
\bibitem[Abed Abud et al.(2021)]{Abed2021} Abed Abud, A., Abi, B., Acciarri, R., et al.\ 2021, arXiv:2103.13910. doi:10.48550/arXiv.2103.13910
\bibitem[An et al.(2016)]{An2016} An, F., An, G., An, Q., et al.\ 2016, Journal of Physics G Nuclear Physics, 43, 030401. doi:10.1088/0954-3899/43/3/030401
\bibitem[Anandagoda et al.(2023)]{Anandagoda2023} Anandagoda, S., Hartmann, D.~H., Fryer, C.~L., et al.\ 2023, \apj, 950, 29. doi:10.3847/1538-4357/acc84f
\bibitem[Ando \& Sato(2004)]{Ando2004} Ando, S. \& Sato, K.\ 2004, New Journal of Physics, 6, 170. doi:10.1088/1367-2630/6/1/170
\bibitem[Ashida \& Nakazato(2022)]{Ashida2022} Ashida, Y. \& Nakazato, K.\ 2022, \apj, 937, 30. doi:10.3847/1538-4357/ac8a46
\bibitem[Ashida et al.(2023)]{Ashida2023} Ashida, Y., Nakazato, K., \& Tsujimoto, T.\ 2023, \apj, 953, 151. doi:10.3847/1538-4357/ace3ba
\bibitem[Bardeen et al.(1972)]{Bardeen1972} Bardeen, J.~M., Press, W.~H., \& Teukolsky, S.~A.\ 1972, \apj, 178, 347. doi:10.1086/151796
\bibitem[Beacom(2010)]{Beacom2010} Beacom, J.~F.\ 2010, Annual Review of Nuclear and Particle Science, 60, 439. doi:10.1146/annurev.nucl.010909.083331
\bibitem[Bisnovatyi-Kogan \& Lamzin(1984)]{Bisnovatyi1984} Bisnovatyi-Kogan, G.~S. \& Lamzin, S.~A.\ 1984, \sovast, 28, 187
\bibitem[Blandford \& Znajek(1977)]{Blandford1977} Blandford, R.~D. \& Znajek, R.~L.\ 1977, \mnras, 179, 433. doi:10.1093/mnras/179.3.433
\bibitem[Branch \& Wheeler(2017)]{Branch2017} Branch, D. \& Wheeler, J.~C.\ 2017, Supernova Explosions: Astronomy and Astrophysics Library, Springer-Verlag (GmbH Germany), 2017. doi:10.1007/978-3-662-55054-0
\bibitem[Burrows et al.(2020)]{Burrows2020} Burrows, A., Radice, D., Vartanyan, D., et al.\ 2020, \mnras, 491, 2715. doi:10.1093/mnras/stz3223
\bibitem[Caballero et al.(2012)]{Caballero2012} Caballero, O.~L., McLaughlin, G.~C., \& Surman, R.\ 2012, \apj, 745, 170. doi:10.1088/0004-637X/745/2/170
\bibitem[Caballero et al.(2016)]{Caballero2016} Caballero, O.~L., Zielinski, T., McLaughlin, G.~C., et al.\ 2016, \prd, 93, 123015. doi:10.1103/PhysRevD.93.123015
\bibitem[Carter(1968)]{Carter1968} Carter, B.\ 1968, Physical Review, 174, 1559. doi:10.1103/PhysRev.174.1559
\bibitem[Chan et al.(2020)]{Chan2020} Chan, C., M{\"u}ller, B., \& Heger, A.\ 2020, \mnras, 495, 3751. doi:10.1093/mnras/staa1431
\bibitem[Chen \& Beloborodov(2007)]{Chen2007} Chen, W.-X. \& Beloborodov, A.~M.\ 2007, \apj, 657, 383. doi:10.1086/508923
\bibitem[Chevalier(1989)]{Chevalier1989} Chevalier, R.~A.\ 1989, \apj, 346, 847. doi:10.1086/168066
\bibitem[de Gouv{\^e}a et al.(2022)]{Gouvea2022} de Gouv{\^e}a, A., Martinez-Soler, I., Perez-Gonzalez, Y.~F., et al.\ 2022, \prd, 106, 103026. doi:10.1103/PhysRevD.106.103026
\bibitem[Dexter \& Kasen(2013)]{Dexter2013} Dexter, J. \& Kasen, D.\ 2013, \apj, 772, 30. doi:10.1088/0004-637X/772/1/30
\bibitem[Ekanger et al.(2022)]{Ekanger2022} Ekanger, N., Horiuchi, S., Kotake, K., et al.\ 2022, \prd, 106, 043026. doi:10.1103/PhysRevD.106.043026
\bibitem[Ekanger et al.(2023)]{Ekanger2023} Ekanger, N., Horiuchi, S., Nagakura, H., et al.\ 2023, arXiv:2310.15254. doi:10.48550/arXiv.2310.15254
\bibitem[Fanton et al.(1997)]{Fanton1997} Fanton, C., Calvani, M., de Felice, F., et al.\ 1997, \pasj, 49, 159. doi:10.1093/pasj/49.2.159
\bibitem[Fischer et al.(2009)]{Fischer2009} Fischer, T., Whitehouse, S.~C., Mezzacappa, A., et al.\ 2009, \aap, 499, 1. doi:10.1051/0004-6361/200811055
\bibitem[Fryer(1999)]{Fryer1999} Fryer, C.~L.\ 1999, \apj, 522, 413. doi:10.1086/307647
\bibitem[Fryer(2006)]{Fryer2006} Fryer, C.~L.\ 2006, \nar, 50, 492. doi:10.1016/j.newar.2006.06.052
\bibitem[Fryer(2009)]{Fryer2009} Fryer, C.~L.\ 2009, \apj, 699, 409. doi:10.1088/0004-637X/699/1/409
\bibitem[Galais et al.(2010)]{Galais2010} Galais, S., Kneller, J., Volpe, C., et al.\ 2010, \prd, 81, 053002. doi:10.1103/PhysRevD.81.053002
\bibitem[Gu(2015)]{Gu2015} Gu, W.-M.\ 2015, \apj, 799, 71. doi:10.1088/0004-637X/799/1/71
\bibitem[Gu et al.(2006)]{Gu2006} Gu, W.-M., Liu, T., \& Lu, J.-F.\ 2006, \apjl, 643, L87. doi:10.1086/505140
\bibitem[Heger \& Woosley(2010)]{Heger2010} Heger, A. \& Woosley, S.~E.\ 2010, \apj, 724, 341. doi:10.1088/0004-637X/724/1/341
\bibitem[Hopkins \& Beacom(2006)]{Hopkins2006} Hopkins, A.~M. \& Beacom, J.~F.\ 2006, \apj, 651, 142. doi:10.1086/506610
\bibitem[Horiuchi et al.(2009)]{Horiuchi2009} Horiuchi, S., Beacom, J.~F., \& Dwek, E.\ 2009, \prd, 79, 083013. doi:10.1103/PhysRevD.79.083013
\bibitem[Horiuchi et al.(2021)]{Horiuchi2021} Horiuchi, S., Kinugawa, T., Takiwaki, T., et al.\ 2021, \prd, 103, 043003. doi:10.1103/PhysRevD.103.043003
\bibitem[Horiuchi et al.(2018)]{Horiuchi2018} Horiuchi, S., Sumiyoshi, K., Nakamura, K., et al.\ 2018, \mnras, 475, 1363. doi:10.1093/mnras/stx3271
\bibitem[Hou et al.(2014)]{Hou2014} Hou, S.-J., Liu, T., Gu, W.-M., et al.\ 2014, \apjl, 781, L19. doi:10.1088/2041-8205/781/1/L19
\bibitem[Janiuk et al.(2007)]{Janiuk2007} Janiuk, A., Yuan, Y., Perna, R., et al.\ 2007, \apj, 664, 1011. doi:10.1086/518761
\bibitem[Kato et al.(2008)]{Kato2008} Kato, S., Fukue, J., \& Mineshige, S.\ 2008, Black-Hole Accretion Disks: Towards a New Paradigm, Kyoto University Press (Kyoto, Japan).
\bibitem[Kawanaka \& Mineshige(2007)]{Kawanaka2007} Kawanaka, N. \& Mineshige, S.\ 2007, \apj, 662, 1156. doi:10.1086/517985
\bibitem[Keil et al.(2003)]{Keil2003} Keil, M.~T., Raffelt, G.~G., \& Janka, H.-T.\ 2003, \apj, 590, 971. doi:10.1086/375130
\bibitem[Kennicutt(1998)]{Kennicutt1998} Kennicutt, R.~C.\ 1998, \araa, 36, 189. doi:10.1146/annurev.astro.36.1.189
\bibitem[Kohri \& Mineshige(2002)]{Kohri2002} Kohri, K. \& Mineshige, S.\ 2002, \apj, 577, 311. doi:10.1086/342166
\bibitem[Kotake et al.(2012)]{Kotake2012} Kotake, K., Takiwaki, T., \& Harikae, S.\ 2012, \apj, 755, 84. doi:10.1088/0004-637X/755/2/84
\bibitem[Kresse et al.(2021)]{Kresse2021} Kresse, D., Ertl, T., \& Janka, H.-T.\ 2021, \apj, 909, 169. doi:10.3847/1538-4357/abd54e
\bibitem[Lee \& Kim(2000)]{Lee2000a} Lee, H.~K. \& Kim, H.-K.\ 2000, Journal of Korean Physical Society, 36, 188. doi:10.48550/arXiv.astro-ph/0008360
\bibitem[Lee et al.(2000)]{Lee2000b} Lee, H.~K., Wijers, R.~A.~M.~J., \& Brown, G.~E.\ 2000, \physrep, 325, 83. doi:10.1016/S0370-1573(99)00084-8
\bibitem[Lee et al.(2005)]{Lee2005} Lee, W.~H., Ramirez-Ruiz, E., \& Page, D.\ 2005, \apj, 632, 421. doi:10.1086/432373
\bibitem[Lei et al.(2009)]{Lei2009} Lei, W.~H., Wang, D.~X., Zhang, L., et al.\ 2009, \apj, 700, 1970. doi:10.1088/0004-637X/700/2/1970
\bibitem[Li et al.(2005)]{Li2005} Li, L.-X., Zimmerman, E.~R., Narayan, R., et al.\ 2005, \apjs, 157, 335. doi:10.1086/428089
\bibitem[Li et al.(2021)]{Li2021} Li, S.~W., Roberts, L.~F., \& Beacom, J.~F.\ 2021, \prd, 103, 023016. doi:10.1103/PhysRevD.103.023016
\bibitem[Li \& Liu(2024)]{Li2024} Li, X.-Y. \& Liu, T.\ 2024, \mnras, 527, 7905. doi:10.1093/mnras/stad3728
\bibitem[Libanov \& Sharofeev(2022)]{Libanov2022} Libanov, A. \& Sharofeev, A.\ 2022, \prd, 106, 123012. doi:10.1103/PhysRevD.106.123012
\bibitem[Liu et al.(2007)]{Liu2007} Liu, T., Gu, W.-M., Xue, L., et al.\ 2007, \apj, 661, 1025. doi:10.1086/513689
\bibitem[Liu et al.(2008)]{Liu2008} Liu, T., Gu, W.-M., Xue, L., et al.\ 2008, \apj, 676, 545. doi:10.1086/527670
\bibitem[Liu et al.(2017)]{Liu2017} Liu, T., Gu, W.-M., \& Zhang, B.\ 2017, \nar, 79, 1. doi:10.1016/j.newar.2017.07.001
\bibitem[Liu et al.(2012)]{Liu2012} Liu, T., Liang, E.-W., Gu, W.-M., et al.\ 2012, \apj, 760, 63. doi:10.1088/0004-637X/760/1/63
%\bibitem[Liu et al.(2015)]{Liu2015} Liu, T., Lin, Y.-Q., Hou, S.-J., et al.\ 2015, \apj, 806, 58. doi:10.1088/0004-637X/806/1/58
\bibitem[Liu et al.(2021b)]{Liu2021b} Liu, T., Qi, Y.-Q., Cai, Z.-Y., et al.\ 2021b, \apj, 920, 5. doi:10.3847/1538-4357/ac1428
\bibitem[Liu et al.(2018)]{Liu2018} Liu, T., Song, C.-Y., Zhang, B., et al.\ 2018, \apj, 852, 20. doi:10.3847/1538-4357/aa9e4f
\bibitem[Liu et al.(2021a)]{Liu2021} Liu, T., Wei, Y.-F., Xue, L., et al.\ 2021a, \apj, 908, 106. doi:10.3847/1538-4357/abd24e
\bibitem[Liu et al.(2013)]{Liu2013} Liu, T., Xue, L., Gu, W.-M., et al.\ 2013, \apj, 762, 102. doi:10.1088/0004-637X/762/2/102
\bibitem[Liu et al.(2014)]{Liu2014} Liu, T., Yu, X.-F., Gu, W.-M., et al.\ 2014, \apj, 791, 69. doi:10.1088/0004-637X/791/1/69
\bibitem[Liu et al.(2016)]{Liu2016} Liu, T., Zhang, B., Li, Y., et al.\ 2016, \prd, 93, 123004. doi:10.1103/PhysRevD.93.123004
\bibitem[Lunardini(2009)]{Lunardini2009} Lunardini, C.\ 2009, \prl, 102, 231101. doi:10.1103/PhysRevLett.102.231101
\bibitem[Lunardini(2016)]{Lunardini2016} Lunardini, C.\ 2016, Astroparticle Physics, 79, 49. doi:10.1016/j.astropartphys.2016.02.005
\bibitem[MacFadyen \& Woosley(1999)]{MacFadyen1999} MacFadyen, A.~I. \& Woosley, S.~E.\ 1999, \apj, 524, 262. doi:10.1086/307790
\bibitem[MacFadyen et al.(2001)]{MacFadyen2001} MacFadyen, A.~I., Woosley, S.~E., \& Heger, A.\ 2001, \apj, 550, 410. doi:10.1086/319698
\bibitem[Maiolino et al.(2008)]{Maiolino2008} Maiolino, R., Nagao, T., Grazian, A., et al.\ 2008, \aap, 488, 463. doi:10.1051/0004-6361:200809678
\bibitem[McLaughlin \& Surman(2007)]{McLaughlin2007} McLaughlin, G.~C. \& Surman, R.\ 2007, \prd, 75, 023005. doi:10.1103/PhysRevD.75.023005
\bibitem[Moriya et al.(2019)]{Moriya2019} Moriya, T.~J., M{\"u}ller, B., Chan, C., et al.\ 2019, \apj, 880, 21. doi:10.3847/1538-4357/ab2643
\bibitem[Moriya et al.(2010)]{Moriya2010} Moriya, T., Tominaga, N., Tanaka, M., et al.\ 2010, \apj, 719, 1445. doi:10.1088/0004-637X/719/2/1445
\bibitem[M{\o}ller et al.(2018)]{Muller2018} M{\o}ller, K., Suliga, A.~M., Tamborra, I., et al.\ 2018, \jcap, 2018, 066. doi:10.1088/1475-7516/2018/05/066
\bibitem[Nagataki \& Kohri(2002)]{Nagataki2002} Nagataki, S. \& Kohri, K.\ 2002, Progress of Theoretical Physics, 108, 789. doi:10.1143/PTP.108.789
\bibitem[Nagataki et al.(2003)]{Nagataki2003} Nagataki, S., Kohri, K., Ando, S., et al.\ 2003, Astroparticle Physics, 18, 551. doi:10.1016/S0927-6505(02)00180-9
\bibitem[Nakazato(2013)]{Nakazato2013} Nakazato, K.\ 2013, \prd, 88, 083012. doi:10.1103/PhysRevD.88.083012
\bibitem[Nakazato et al.(2015)]{Nakazato2015} Nakazato, K., Mochida, E., Niino, Y., et al.\ 2015, \apj, 804, 75. doi:10.1088/0004-637X/804/1/75
\bibitem[Nakazato et al.(2008)]{Nakazato2008} Nakazato, K., Sumiyoshi, K., Suzuki, H., et al.\ 2008, \prd, 78, 083014. doi:10.1103/PhysRevD.78.083014
\bibitem[Narayan et al.(2001)]{Narayan2001} Narayan, R., Piran, T., \& Kumar, P.\ 2001, \apj, 557, 949. doi:10.1086/322267
\bibitem[O'Connor \& Ott(2011)]{OConnor2011} O'Connor, E. \& Ott, C.~D.\ 2011, \apj, 730, 70. doi:10.1088/0004-637X/730/2/70
\bibitem[Paxton et al.(2011)]{Paxton2011} Paxton, B., Bildsten, L., Dotter, A., et al.\ 2011, \apjs, 192, 3. doi:10.1088/0067-0049/192/1/3
\bibitem[Paxton et al.(2013)]{Paxton2013} Paxton, B., Cantiello, M., Arras, P., et al.\ 2013, \apjs, 208, 4. doi:10.1088/0067-0049/208/1/4
\bibitem[Peeples \& Somerville(2013)]{Peeples2013} Peeples, M.~S. \& Somerville, R.~S.\ 2013, \mnras, 428, 1766. doi:10.1093/mnras/sts158
\bibitem[Perna et al.(2014)]{Perna2014} Perna, R., Duffell, P., Cantiello, M., et al.\ 2014, \apj, 781, 119. doi:10.1088/0004-637X/781/2/119
\bibitem[Popham et al.(1999)]{Popham1999} Popham, R., Woosley, S.~E., \& Fryer, C.\ 1999, \apj, 518, 356. doi:10.1086/307259
\bibitem[Qi et al.(2022b)]{Qi2022b} Qi, Y.-Q., Liu, T., Cai, Z.-Y., et al.\ 2022b, \apj, 934, 1. doi:10.3847/1538-4357/ac7a43
\bibitem[Qi et al.(2022a)]{Qi2022} Qi, Y.-Q., Liu, T., Huang, B.-Q., et al.\ 2022a, \apj, 925, 43. doi:10.3847/1538-4357/ac3757
\bibitem[Qu \& Liu(2022)]{Qu2022} Qu, H.-M. \& Liu, T.\ 2022, \apj, 929, 83. doi:10.3847/1538-4357/ac5f4b
\bibitem[Rauch \& Blandford(1994)]{Rauch1994} Rauch, K.~P. \& Blandford, R.~D.\ 1994, \apj, 421, 46. doi:10.1086/173625
\bibitem[Reddy et al.(2008)]{Reddy2008} Reddy, N.~A., Steidel, C.~C., Pettini, M., et al.\ 2008, \apjs, 175, 48. doi:10.1086/521105
\bibitem[Rujopakarn et al.(2010)]{Rujopakarn2010} Rujopakarn, W., Eisenstein, D.~J., Rieke, G.~H., et al.\ 2010, \apj, 718, 1171. doi:10.1088/0004-637X/718/2/1171
\bibitem[Salpeter(1955)]{Salpeter1955} Salpeter, E.~E.\ 1955, \apj, 121, 161. doi:10.1086/145971
\bibitem[Schilbach et al.(2019)]{Schilbach2019} Schilbach, T.~S.~H., Caballero, O.~L., \& McLaughlin, G.~C.\ 2019, \prd, 100, 043008. doi:10.1103/PhysRevD.100.043008
\bibitem[Sekiguchi et al.(2011)]{Sekiguchi2011} Sekiguchi, Y., Kiuchi, K., Kyutoku, K., et al.\ 2011, \prl, 107, 211101. doi:10.1103/PhysRevLett.107.211101
\bibitem[Song \& Liu(2019)]{Song2019} Song, C.-Y. \& Liu, T.\ 2019, \apj, 871, 117. doi:10.3847/1538-4357/aaf6ae
\bibitem[Song et al.(2015)]{Song2015} Song, C.-Y., Liu, T., Gu, W.-M., et al.\ 2015, \apj, 815, 54. doi:10.1088/0004-637X/815/1/54
\bibitem[Song et al.(2016)]{Song2016} Song, C.-Y., Liu, T., Gu, W.-M., et al.\ 2016, \mnras, 458, 1921. doi:10.1093/mnras/stw427
\bibitem[Song et al.(2020)]{Song2020} Song, C.-Y., Liu, T., \& Wei, Y.-F.\ 2020, \mnras, 494, 3962. doi:10.1093/mnras/staa932
\bibitem[Strumia \& Vissani(2003)]{Strumia2003} Strumia, A. \& Vissani, F.\ 2003, Physics Letters B, 564, 42. doi:10.1016/S0370-2693(03)00616-6
\bibitem[Sukhbold \& Woosley(2014)]{Sukhbold2014} Sukhbold, T. \& Woosley, S.~E.\ 2014, \apj, 783, 10. doi:10.1088/0004-637X/783/1/10
\bibitem[Suliga(2022)]{Suliga2022} Suliga, A.~M.\ 2022, arXiv:2207.09632. doi:10.48550/arXiv.2207.09632
\bibitem[Sumiyoshi et al.(2007)]{Sumiyoshi2007} Sumiyoshi, K., Yamada, S., \& Suzuki, H.\ 2007, \apj, 667, 382. doi:10.1086/520876
\bibitem[Sumiyoshi et al.(2008)]{Sumiyoshi2008} Sumiyoshi, K., Yamada, S., \& Suzuki, H.\ 2008, \apj, 688, 1176. doi:10.1086/592183
\bibitem[Sumiyoshi et al.(2006)]{Sumiyoshi2006} Sumiyoshi, K., Yamada, S., Suzuki, H., et al.\ 2006, \prl, 97, 091101. doi:10.1103/PhysRevLett.97.091101
\bibitem[Tabrizi \& Horiuchi(2021)]{Tabrizi2021} Tabrizi, Z. \& Horiuchi, S.\ 2021, \jcap, 2021, 011. doi:10.1088/1475-7516/2021/05/011
\bibitem[Vitagliano et al.(2020)]{Vitagliano2020} Vitagliano, E., Tamborra, I., \& Raffelt, G.\ 2020, Reviews of Modern Physics, 92, 045006. doi:10.1103/RevModPhys.92.045006
\bibitem[Vogel \& Beacom(1999)]{Vogel1999} Vogel, P. \& Beacom, J.~F.\ 1999, \prd, 60, 053003. doi:10.1103/PhysRevD.60.053003
\bibitem[Weaver et al.(1978)]{Weaver1978} Weaver, T.~A., Zimmerman, G.~B., \& Woosley, S.~E.\ 1978, \apj, 225, 1021. doi:10.1086/156569
\bibitem[Wei \& Liu(2022)]{Wei2022} Wei, Y.-F. \& Liu, T.\ 2022, Universe, 8, 529. doi:10.3390/universe8100529
\bibitem[Wei et al.(2019)]{Wei2019} Wei, Y.-F., Liu, T., \& Song, C.-Y.\ 2019, \apj, 878, 142. doi:10.3847/1538-4357/ab2187
\bibitem[Wei et al.(2021)]{Wei2021} Wei, Y.-F., Liu, T., \& Xue, L.\ 2021, \mnras, 507, 431. doi:10.1093/mnras/stab2153
\bibitem[White et al.(2016)]{White2016} White, C.~J., Stone, J.~M., \& Gammie, C.~F.\ 2016, \apjs, 225, 22. doi:10.3847/0067-0049/225/2/22
\bibitem[Wong et al.(2014)]{Wong2014} Wong, T.-W., Fryer, C.~L., Ellinger, C.~I., et al.\ 2014, arXiv:1401.3032. doi:10.48550/arXiv.1401.3032
\bibitem[Woosley(1989)]{Woosley1989} Woosley, S.~E.\ 1989, Annals of the New York Academy of Sciences, 571, 397. doi:10.1111/j.1749-6632.1989.tb50526.x
\bibitem[Woosley(1993)]{Woosley1993} Woosley, S.~E.\ 1993, \apj, 405, 273. doi:10.1086/172359
\bibitem[Woosley \& Heger(2007)]{Woosley2007} Woosley, S.~E. \& Heger, A.\ 2007, \physrep, 442, 269. doi:10.1016/j.physrep.2007.02.009
\bibitem[Woosley et al.(2002)]{Woosley2002} Woosley, S.~E., Heger, A., \& Weaver, T.~A.\ 2002, Reviews of Modern Physics, 74, 1015. doi:10.1103/RevModPhys.74.1015
\bibitem[Woosley \& Weaver(1995)]{Woosley1995} Woosley, S.~E. \& Weaver, T.~A.\ 1995, \apjs, 101, 181. doi:10.1086/192237
\bibitem[Wu et al.(2013)]{Wu2013} Wu, X.-F., Hou, S.-J., \& Lei, W.-H.\ 2013, \apjl, 767, L36. doi:10.1088/2041-8205/767/2/L36
\bibitem[Xue et al.(2013)]{Xue2013} Xue, L., Liu, T., Gu, W.-M., et al.\ 2013, \apjs, 207, 23. doi:10.1088/0067-0049/207/2/23
\bibitem[Y{\"u}ksel \& Kistler(2015)]{Yuksel2015} Y{\"u}ksel, H. \& Kistler, M.~D.\ 2015, Physics Letters B, 751, 413. doi:10.1016/j.physletb.2015.10.055
\bibitem[Y{\"u}ksel et al.(2008)]{Yuksel2008} Y{\"u}ksel, H., Kistler, M.~D., Beacom, J.~F., et al.\ 2008, \apjl, 683, L5. doi:10.1086/591449
\bibitem[Zhang et al.(2008)]{Zhang2008} Zhang, W., Woosley, S.~E., \& Heger, A.\ 2008, \apj, 679, 639. doi:10.1086/526404
\end{thebibliography}
\end{document}